\documentclass[reprint,amsmath,amssymb,aps,prb,groupedaddress,nofootinbib,superscriptaddress]{revtex4-1}
\usepackage{graphicx,xcolor,tikz}
\usepackage{amsthm,amssymb,amsmath,dsfont,braket}
\usepackage{bm}
\usepackage[hidelinks,pagebackref=false,pdfnewwindow=true]{hyperref} 
\usepackage{epstopdf,psfrag}
\usepackage{relsize,amsbsy}

\newcommand{\mac}{\mathcal}
\newcommand{\be}{\begin{equation}}
\newcommand{\ee}{\end{equation}}

\newcommand{\1}{\mathds{1}}
\newcommand{\ms}{Majorana spinon }
\newcommand{\mss}{Majorana spinons }
\newcommand{\mssns}{Majorana spinons}
\newcommand{\hh}{harmonic honeycomb }
\newcommand{\HH}{Harmonic Honeycomb }
\newcommand{\Hs}[1]{\mbox{$\mathcal{H}$-$#1$}}

\newcommand{\bit}{\begin{enumerate}}
\newcommand{\eit}{\end{enumerate}}
\newcommand{\m}{\item}
\newcommand{\zt}{\mathbb{Z}_2}

\definecolor{bananayellow}{rgb}{1.0, 0.88, 0.21}
\definecolor{straw}{rgb}{0.75, 0.71, 0.3}

\definecolor{palatinatepurple}{rgb}{0.49, 0.24, 0.46}

\definecolor{darkblue}{rgb}{0.0, 0.0, 0.55}

\definecolor{darkgreen}{rgb}{0.0, 0.5, 0.0}

\begin{document}

\title{Theory of Raman response in three-dimensional Kitaev spin liquids:\\ application to $\beta-$ and $\gamma-$Li$_2$IrO$_3$ compounds}
		\author{Brent Perreault}
	\affiliation{\small School of Physics and Astronomy, University of Minnesota, Minneapolis, Minnesota 55455, USA}
%
	\author{Johannes Knolle}
	\affiliation{\small Department of Physics, Cavendish Laboratory, JJ Thomson Avenue, Cambridge CB3 0HE, U.K.}
	\author{Natalia B. Perkins}
	\author{F. J. Burnell}
	\affiliation{\small School of Physics and Astronomy, University of Minnesota, Minneapolis, Minnesota 55455, USA}
	
	\date{\today}

	\begin{abstract}
	We calculate the Raman response for the Kitaev spin model on the $\mathcal{H}$-$0$, $\mathcal{H}$-$1$, and $\mathcal{H}$-$\infty$ harmonic honeycomb lattices. We identify several quantitative features in the Raman spectrum that are characteristic of the spin liquid phase. Unlike the dynamical structure factor, which probes both the Majorana spinons and flux excitations that emerge from spin fractionalization, the Raman spectrum in the Kitaev models directly probes a density of states of pairs of fractional, dispersing Majorana spinons. As a consequence, the Raman spectrum in all these models is gapless for sufficiently isotropic couplings, with a low-energy power law that results from the Fermi lines (or points) of the dispersing Majorana spinons. We show that the polarization dependence of the Raman spectrum contains  crucial information about the symmetry of the ground state. We also discuss to what extent the features of the Raman response that we find reflect generic properties of the spin liquid phase, and comment on their possible relevance to $\alpha-$, $\beta-$, and $\gamma-$Li$_2$IrO$_3$ compounds. 	
	\end{abstract}

	
	\maketitle
	

	\section{Introduction}

	Quantum spin liquids (QSLs) have been a topic of significant interest for four decades since they were introduced in the seminal work of Anderson.\cite{Anderson}  
	The search for these exotic phases of matter has focused on low-dimensional and highly frustrated magnets, where conventional ordering is suppressed and quantum fluctuations are large, opening up the possibility of a ground state that preserves all the symmetries of the underlying spin Hamiltonian. 
	In lieu of a conventional magnetic order associated with broken symmetry, these symmetry-preserving QSL phases are characterised by a combination of properties known as {\it topological order}.\cite{Wen02,Balents,Moessner01}  
	
	Recent years have seen remarkable progress both in the theoretical understanding of the nature of QSLs and in identifying realistic Heisenberg models which might host them. \cite{Yan2011,Jiang2012,Iqbal2011,Meng2010,Depenbrock2012,Gong2015,Lawler08} 
	In parallel, a long experimental quest has identified a number of materials as spin liquid candidates, and evidence for spin liquid physics exists in several two- and three-dimensional geometrically frustrated magnets, including  triangular, \cite{Yamashita08,Shimizu03,Itou08,Yamashita10, Zhou11} 		
	kagome,\cite{Fak2012}  hyperkagome\cite{Okamoto07}  and pyrochlore antiferromagnets.\cite{Tokiwa2014,Sibille15} 
	
	In spite of this progress, however, an unambiguous identification of QSL phases in these systems remains elusive.
	The difficulty in experimentally identifying QSLs comes from the fact that, unlike states with broken symmetry, the {\it topological order}  characteristic of QSLs does not admit a local order parameter. 
	This makes it challenging to identify ``smoking gun" experimental signatures.

However, one distinctive feature of topological order that can be probed by conventional methods,\cite{Hermele} such as inelastic neutron\cite{Coldea01,Coldea,Helton07,Helton10,Han,Fak2012} or Raman scattering\cite{Lemmens03,Wulferding,wulferding2}, is fractionalization. In quantum spin liquids there
are elementary (chargeless) excitations carrying fractional quantum numbers
relative to the local constituent degrees of freedom
(charged electrons and spin 1 magnons). Therefore only multiple
quasiparticles can couple to external scattering probes. With access to the full spectrum, one can in principle disentangle the contributions of individual quasiparticles with different energies from the resulting multi-particle continuum, and make a quantitative comparison between theory and experiment. 
This is in contrast with thermodynamic measurements,\cite{Yamashita08,Helton07,Yamashita09} which only probe the asymptotic low energy response. 

The power of such a quantitative comparison was strikingly demonstrated in the one-dimensional systems CuSO$_4\cdot 5$D$_2$O and ${\mathrm{KCuF}}_{3}$. \cite{Tennant,Mourigal,Muller,Caux2006,Fadeev81} Suggestively, the experiments\cite{Tennant,Mourigal} see broad spectra characteristic of fractionalized excitations. 
The fractionalization in these 1D systems was definitively identified by excellent 
{\it quantitative} agreement between experiments\cite{Tennant,Mourigal} and exact calculations of the two- and four-spinon dynamical structure factor based on an exact solution of the Heisenberg model in one dimension using Bethe Ansatz,\cite{Muller,Caux2006} which shows fractionalized spin excitations.\cite{Fadeev81}

Broad spectra have also been predicted and observed in two- and three-dimensional candidate QSL materials. \cite{Coldea01,Coldea,Lake05,Han,Wulferding,wulferding2,Cepas,Ko,Perkins13}
However, the level of quantitative agreement between experiment and theory demonstrated in 1D has yet to be achieved in higher dimensions.\cite{Banerjee,Knolle} In large part, this is because obtaining a reliable calculation of the spin response functions for spin liquid states in Heisenberg models has proven difficult due to a lack of controlled methods that can be used to treat these highly frustrated systems. For the Heisenberg QSL candidates theoretical calculations are restricted either to numerics, for which obtaining entire spectra in 2D and 3D models is quite challenging (see Ref. \onlinecite{Depenbrock2012} and references therein), or to variational-type ansatz such as slave particle\cite{Punk,Messio10,Dodds,Mezio11,Merino} or resonant-valence-bond treatments.\cite{Hao09,Hao10,Hao13}

However, for a particular class of spin-exchange Hamiltonians based on the Kitaev honeycomb model,\cite{Kitaev} the theoretical situation is much more tractable. For these (integrable) models both the dynamical structure factor\cite{Knolle} and the Raman response\cite{Perkins} can be obtained analytically. This is highly significant, since any quantitative features of the resulting spectrum that are characteristic of the spin-liquid phase can
provide a reliable basis for comparison with experiments.

In this context, an exciting current development in the search for spin liquids has been the synthesis of transition metal compounds belonging to the A$_2$IrO$_3$ iridate family, \cite{Singh,Singh2,Choi,Biffin,Biffin2,Takayama} and more recently, $\alpha$-RuCl$_3$. \cite{Plumb,Sears}  These materials form trivalent lattice structures in the {\it \hh } family and have magnetic moments arising from Ir$^{4+}$ or Ru$^{3+}$ ions, which due to strong spin-orbit coupling are characterized by $J_{eff}=1/2$ states. Edge-sharing IrO$_6$ or RuCl$_6$ octahedra provide 90$^{\circ}$ paths for a Kitaev-like super-exchange coupling among magnetic moments.\cite{Jackeli,Chaloupka,Katukuri,Sizyuk,Shankar2}
Although the known candidate compounds A$_2$IrO$_3$, as well as  $\alpha$-RuCl$_3$, actually order at low temperature, \cite{Singh} the presence of a large Kitaev term suggests that these ordered ground states are proximate to spin liquid phases -- a statement which is also supported by recent experiments. \cite{Biffin,Biffin,Gupta,Chun15,Sandilands,Banerjee}
	
The tractability of the Kitaev model allows for explicit  computations of the dynamics.\cite{Knolle,Perkins} 
 In particular, spins fractionalize into two types of degrees of freedom in the Kitaev model: dispersing \mss and gapped flux loops. In the unperturbed Kitaev model the flux is conserved (static), and the ground state can be viewed as a band insulator, or metal, of \mss in the flux background which minimizes the total energy. The dynamical structure factor exhibits a gap that can be understood as the energy to excite a pair of fluxes (a flux loop in 3D) because spin excitations fractionalize into both dispersing \mss and gapped fluxes. 

These predictions for the 2D model have already been tested experimentally and can be understood within the fractionalization of the Kitaev model.  
While the standard tool for measuring the dynamical structure factor -- neutron scattering -- is challenging in iridates because of strong neutron absorption, neutron scattering data have been recently reported for $\alpha$-RuCl$_3$.\cite{Banerjee}  Both a gapped spectrum and a broad continuum were observed and are consistent with the theoretical calculation.\cite{Knolle}. On the other hand, low-energy photons create pairs of \mss (no fluxes), allowing the Raman operator to directly probe a two-\ms density of states (DOS).
Indeed, Raman scattering  experiments in $\alpha$-RuCl$_3$ \cite{Sandilands} 
show a broad scattering continuum consistent with this prediction, as well as the prediction of no dependence on polarization.\cite{Perkins, Cepas} Similar but somewhat controversial Raman scattering data have been reported for Na$_2$IrO$_3$.\cite{Gupta}

		\begin{figure*}
			\includegraphics[width=0.3\linewidth]{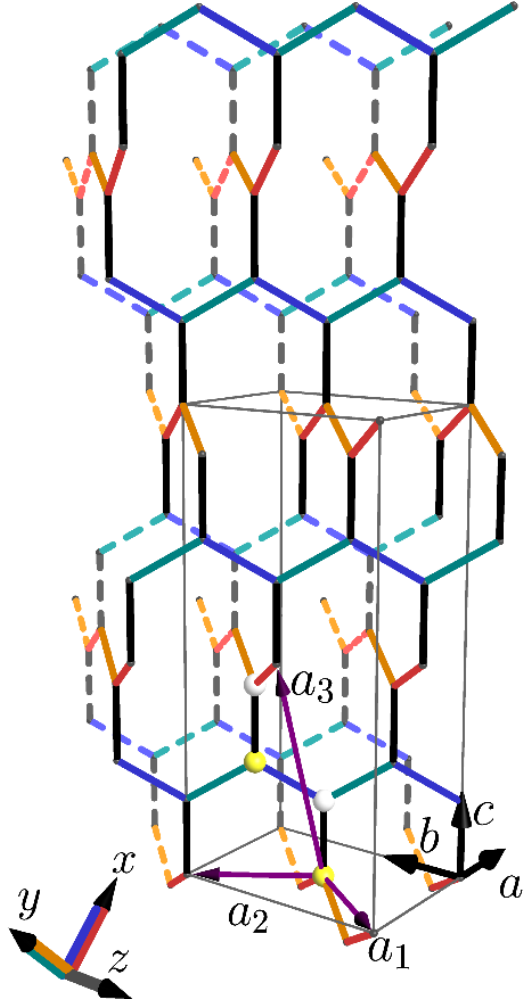}
			\includegraphics[width=0.31\linewidth]{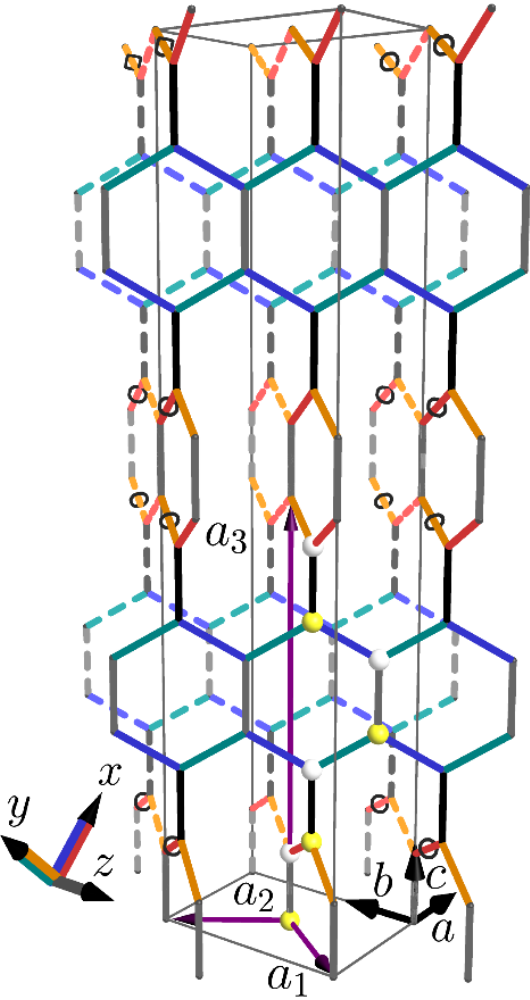}	
			\begin{tikzpicture}
			\node[anchor=south west,inner sep=0] (image) at (0,0) {
				\includegraphics[width=0.26\linewidth]{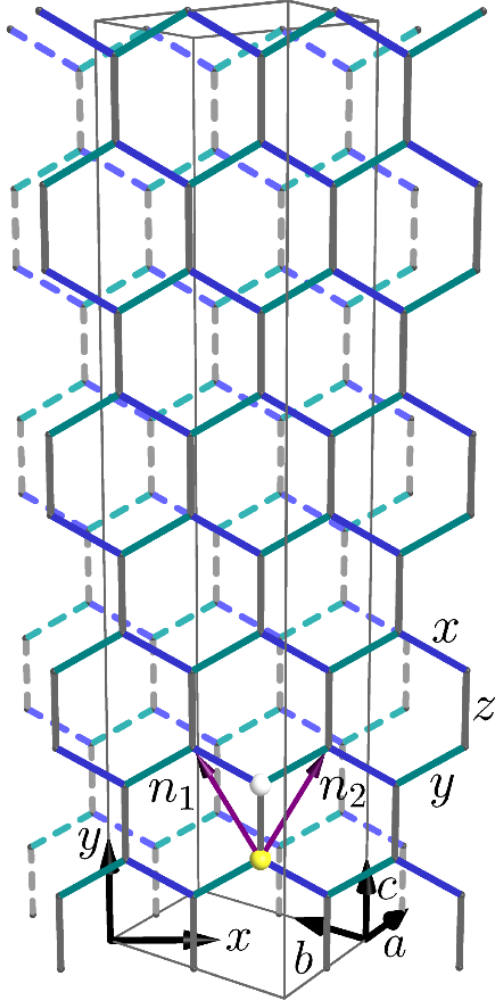} };
			\node[anchor=south west,inner sep=0] (image) at (-1,4.8) {\Huge ...} ;
			\end{tikzpicture}	
	\caption{(Color online) The \hh (\Hs{n}) lattices: \Hs{0} (left), \Hs{1} (center), \Hs{\infty} (right).  Here $n$ counts the number of rows along the $c$-axis before the orientation of the honeycomb plane switches between the two non-parallel chains of $x$ and $y$ bonds. The \Hs{0} Hyperhoneycomb lattice switches chains at every $c$-bond. The \Hs{1} has one set of $c$-bonds making rungs (gray) of 
ladders before a bridge ($c$-bonds in  black) to the opposite ladder. The \Hs{\infty} never switches ladders. The ladders are labeled  by red(orange) and blue(green)  on $x$($y$)-bonds. The primitive unit cell for $n<\infty$ contains $4n+4$ sites. The honeycomb lattice has only two sites in the unit cell. Our choice of unit cell is illustrated by the spheres, whose color alternates yellow and white indicating the odd and even sublattices respectively. The excited bonds in the $\pi$ flux state are circled on the \Hs{1} lattice.}
\label{fig:Harmonic_Lattices}
		\end{figure*}
		
	The experiments on two dimensional compounds illustrate the power of dynamical scattering experiments for identifying these elusive phases of matter. 
	Our work focuses on calculations  of dynamical Raman spectra for the 2D and 3D harmonic honeycomb systems, in particular on the $\Hs{\infty}$, $\Hs{0}$, and $\Hs{1}$ lattices realized in the $\alpha-$, $\beta-$ and $\gamma-$Li$_2$IrO$_3$ compounds.\cite{Modic,Takayama}  	We find that the spin liquid's Raman spectrum displays a broad continuum  related to  the two-\ms DOS (2-DOS).  In the gapless phase there is spectral weight down to zero frequency. In fact, the low energy asymptotic spectrum reflects the \ms Fermi surface topology. In addition, the 3D lattice structures lead to a richer polarization dependence as well as more spectral features than their 2D counterpart, due to the larger unit cells and lower symmetry.
	We show that most of these features result from the high symmetry of the spin liquid phase, together with the particular form of the Raman operator in systems with only  bilinear spin-exchange interactions, and are therefore expected to be characteristic of the QSL phase in these systems.
	
	The structure of the paper is as follows. 
	In section \ref{ModelSec}, we review the Kitaev model on the \hh lattices and give some relevant details of their structures. 
	Section \ref{RamanSec} reviews the theory of Raman scattering in Kitaev spin liquids.  
	Section \ref{ResultsSec} presents the results of our calculations, together with a discussion of their applicability to the  experimentally realizable compounds.  
	A number of technical details are explained in appendices.  
	Appendix \ref{sec:lattice} presents more details of the lattice structures, their symmetries, and the band Hamiltonians of interest. Appendix \ref{Intense} presents some technical details of the derivation of the Raman response. Appendix \ref{pol} gives detailed arguments for the polarization dependence of the Raman signal described in section \ref{PolSec}.

	\section{The model} \label{ModelSec}

	The pure Kitaev model that we work with below is a good starting point for understanding response functions in the spin liquid phase. The advantages of  this model are threefold.  
	First, it is exactly solvable on the whole family of \hh lattices. 
	Second, both gapped and gapless spin liquid ground states can be obtained within this model by tuning the anisotropy of the spin-spin interactions on different bonds.  Finally, the fractionalization  of spin excitations  emerges naturally in the exact solution, \cite{Kitaev} allowing for a clear identification of the role that these fractionalized excitations play in experimental response functions. For example, we show that only one type of fractionalized quasiparticle is probed by the Raman operator, which greatly simplifies the interpretation of the dynamical response.

The Kitaev model\cite{Kitaev} on a generic tri-coordinated lattices takes the form
\begin{align}\label{H}
	\mathcal{H} =
	\sum_{\left<ij\right>^\alpha}  J^\alpha \sigma^\alpha_i \sigma^\alpha_j . 
	\end{align}
    Here $\alpha = x,y$ and $z$  label the three types of bonds emenating from each vertex, and
     $J^x$, $J^y$ and $J^z$ are the associated coupling constants.

\begin{figure}
	\centering
	\includegraphics[width=.55\linewidth, trim = 0mm 15mm 0mm 15mm, clip]{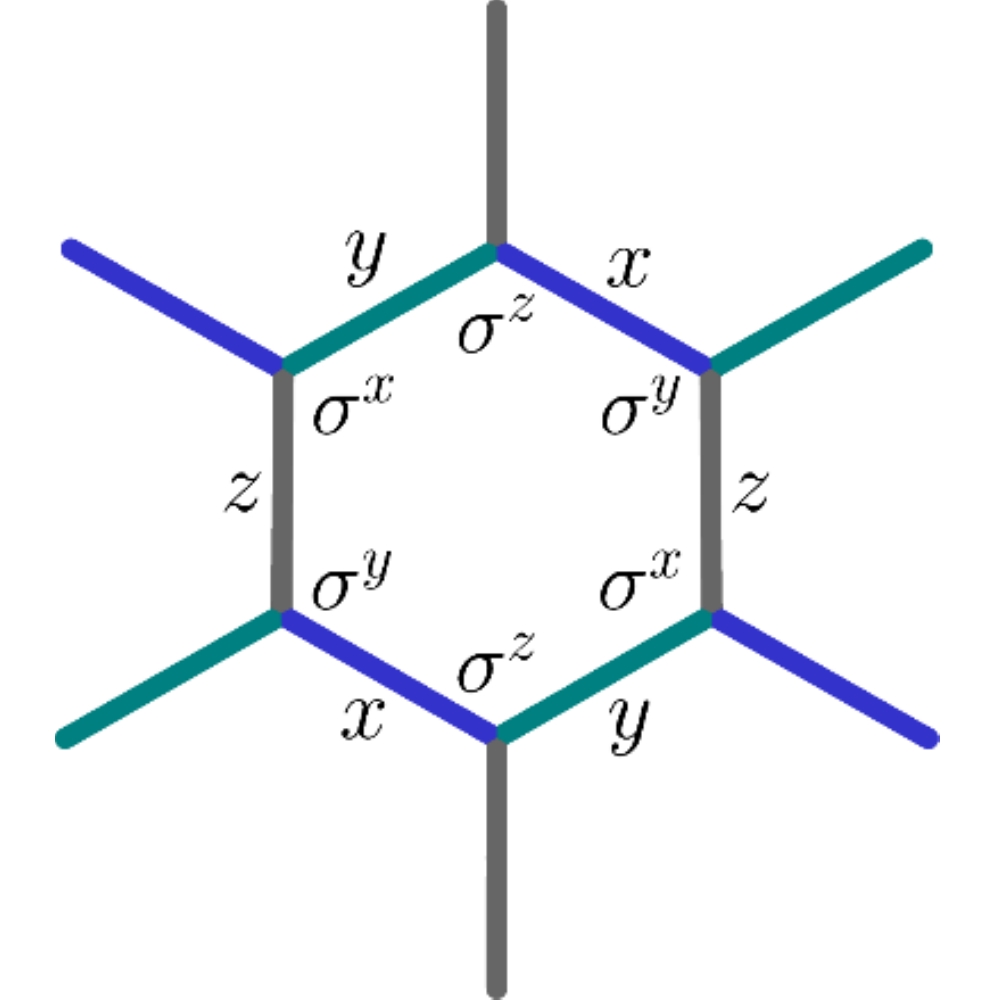}
	\caption{(Color online) An elementary plaquette of the honeycomb lattice with the bonds labeled by the component of spins that interact along them. The spin component shown at each vertex is the one that enters the product $W_P$ of Eq.~(\ref{W}), and corresponds to the color of the bond pointing out of the plaquette from that vertex. 
		}
	\label{fig:ComFig}
\end{figure}

Because only one component of the spin interacts along each bond, there is one conserved quantity for every plaquette.\cite{Kitaev} This conserved quantity is given by
\begin{align}\label{W}
W_P=\prod_{i \in  P} \sigma_i^{\alpha (i)},
\end{align}	
which is the product of spin operators around a  plaquette $P$,  whose spin component $\alpha(i)$ is given by the label of the outgoing bond direction, as illustrated for the honeycomb lattice in Fig. \ref{fig:ComFig}. 

The full spectrum  of the model Eq.~(\ref{H}) can be described by using 
Majorana fermions to represent the spins. Following the notation of the original work by Kitaev, \cite{Kitaev} we use
\begin{align}\label{MarSpins}
	\sigma^\alpha_j = i c_j b^\alpha_j\,.
	\end{align}
 The Majorana fermions  $c_j$ and $b_j^\alpha$ satisfy the algebra
	\begin{align}
	c_j^2 = b_j^2 =1, \hspace{1cm} \{b_j^{\alpha}, b_i^{\beta} \} = \{ b_j^\alpha, c_j \} = 0\,
	\end{align} 
 so that $c^\dagger = c$ and Majorana operators at different sites anticommute.  
	 In terms of Majorana fermions the Hamiltonian becomes
	\begin{align}\label{Hm}
	{H} = i  \sum_{\left<ij\right>^\alpha}J^\alpha u_{\left<ij\right>^\alpha} c_i c_j,
	\end{align}
	where  we have defined the {\it bond operators} 
	\be 
	u_{\left<ij\right>^\alpha} = i b_i^\alpha b_j^\alpha \ ,
	\ee
which satisfy $u_{\left<ij\right>^\alpha} = -u_{\left<ji\right>^\alpha}$.   
Since the bond operator $u_{\left<ij\right>^\alpha}$ commutes with the Hamiltonian in any eigenstate it is a constant. Therefore, in the unconstrained Hilbert space, eigenstates of Eq.~(\ref{Hm}) are described by the value of $u_{\left<ij\right>^\alpha}$ on each edge, together with an eigenstate of the resulting hopping Hamiltonian for the $\{c_j\}$ Majorana fermions.    
Since \mbox{$\left(u_{\left<ij\right>^\alpha}\right)^2 = 1$} then \mbox{$u_{\left<ij\right>^\alpha} = \pm1$} and the possible hopping parameters on each edge are $t_{ij} = \pm i J_{\left<ij\right>^\alpha}$.  

Because the Majorana representation is redundant, not all of the eigenstates described above are physically distinct.   Imposing $1 = D_j = c_j b_j^x b_j^y b_j^z$ at each site makes the representation exact. In particular, though individual $u_{\left<ij\right>^\alpha}$ commute with $H$, they do not commute with the constraint.  
The only conserved quantities corresponding to the $u_{\left<ij\right>^\alpha}$ in the {\it physical} Hilbert space are the $\{ W_P\}$,  defined by Eq.~(\ref {W}), which in the Majorana representation have the form:
\be
W_P = \prod_{\left<ij\right> \in  P} \braket{u_{\left<ij\right>^\alpha}}.
\ee
Thus the physical eigenstates (and the band energies for $\{c_j\}$ Majorana fermions) depend only on the phase accumulated by a fermion hopping around a given plaquette. We define the flux $\Phi_P$ through plaquette $P$ by  
$\exp\left[i\Phi_P\right] = \prod_{\left<ij\right> \in  P} t_{ij}/|t_{ij}| = i^n W_P$, where $n$ is the number of the bonds around  the plaquette $P$. Since $u_{\left<ij\right>^\alpha}= \pm 1$, then $W_P = \pm 1$ and the fluxes -- which effectively generate lattice-scale magnetic flux for the dispersing fermions -- can only take on  two values: $0$ or $\pi$.
From here on we will therefore describe excitations of the $b_j^\alpha$ Majorana operators in terms of their effect on these $\zt$ fluxes, and refer to the dispersing $c_j$ excitations as \textit{\mssns}.

\subsection{The \HH lattices}

Motivated by recent synthesis of  the $\alpha$-, $\beta$- and $\gamma$-Li$_2$IrO$_3$ compounds, here we study the Kitaev model on the \Hs{\infty}, \Hs{0}, and \Hs{1} lattices illustrated in Fig.\ref{fig:Harmonic_Lattices}. The  discussion here generalizes straightforwardly to the other lattices in the harmonic  honeycomb series, \Hs{n}.

 The \hh series consists of bipartite orthorhombic tri-coordinated 3D lattices. The \Hs{\infty} lattice is the exception, splitting into uncoupled 2D honeycomb lattice planes, shown on the right of Fig. \ref{fig:Harmonic_Lattices}. Because the honeycomb lattice is well-known we leave the details of the in-plane coordinates and the unit cell that we use to Appendix \ref{sec:lattice}.   
 The \Hs{n} lattice consists of $n$ rows of co-planar hexagonal plaquettes, followed by a ``bridge"  layer of $\hat{c}$ axis bonds, and then another $n$ rows in a new plane.  
 In terms of the orthorhombic unit vectors ($\hat{\mathbf{a}},\hat{\mathbf{b}},\hat{\mathbf{c}}$ in  Fig. \ref{fig:Harmonic_Lattices}), a natural set of unit vectors (for $n< \infty$) is \cite{Schaffer} 
 \begin{align} \label{LatticeVectorsEq}
	{\bf a}_1 &= (-1,-\sqrt{2},0) \nonumber\\
	{\bf a}_2 &= (-1,\sqrt{2},0) \nonumber\\
	{\bf a}_3 &= \left \{ \begin{array}{ll}
	(-1,0,3) + (0,0,6)\times \frac{n}{2} & \text{if } n \text{ is even,} \\
	(0,0,6)\times \frac{n+1}{2} & \text{if } n \text{ is odd.}
	\end{array}\right.
	\end{align}
where we have set the bond length to $1$. The unit cell therefore contains $4n+4$ sites. 

The spin-orbit coupling distinguished directions ($\hat{\mathbf{x}},\hat{\mathbf{y}},\hat{\mathbf{z}}$ in Fig. \ref{fig:Harmonic_Lattices}), which determine which spin components couple along which bonds,  
 are  $\hat{\mathbf{z}} = \hat{\mathbf{b}}$, $\hat{\mathbf{y}}=(\hat{\mathbf{a}}+\hat{\mathbf{c}})/\sqrt{2}$, and $\hat{\mathbf{x}}=(-\hat{\mathbf{a}}+\hat{\mathbf{c}})/\sqrt{2}$. The nearest-neighbor bond vectors (in the $\hat{\mathbf{a}},\hat{\mathbf{b}},\hat{\mathbf{c}}$ basis) are
	\begin{align}\label{3Dds}
	{\bf d}^z &=    (0,0,1)  &  (\textrm{black and gray})  \nonumber \\
	{\bf d}^{x}_A &=  \frac{1}{2}(1, {\sqrt{2}},-1)  & (\textrm{red}) \nonumber \\ 
	{\bf d}^{x}_B &=  \frac{1}{2}(1, - {\sqrt{2}},-1) &(\textrm{blue})  \nonumber \\ 
	{\bf d}^{y}_A &=  \frac{1}{2}(-1, - {\sqrt{2}},-1)  & (\textrm{orange}) \\ 
	 {\bf d}^{y}_B &=  \frac{1}{2}(-1,  {\sqrt{2}},-1)  & (\textrm{green}) \nonumber,
	\end{align}
	where the bonds are oriented from the odd (yellow) to the even (white) sublattice, and the colors refer to Fig.~\ref{fig:Harmonic_Lattices}. Note that the $x$- and $y$-type couplings occur along bonds of two different orientations, labeled $A$ (red  $x$- and orange $y$-bonds) and $B$ (blue  $x$- and green $y$-bonds). All $z$ bonds are parallel to the $\hat{c}$ axis. 
	
	The spin-exchange Hamiltonian for each lattice is expected to preserve the lattice symmetry, which, due to spin-orbit coupling,
simultaneously acts on the spin and lattice basis vectors.  	
On the 2D honeycomb lattice, $C_6$ rotation symmetry\cite{You} requires that the exchange coupling be the same on each bond: $J^x = J^y = J^z$.  
On the $\mac{H}$-$n$, $n<\infty$ lattices the $C_2$  rotation symmetry  interchanges $J^x$ and $J^y$ bonds, and thus  $J^x=J^y$ .  However,  in  general, $J^z \ne J^x = J^y$, since these are not related by symmetry. 
For comparison with the 3D cases, we will also consider $J^z \ne J^x = J^y$ on the 2D lattice, although this explicitly breaks the $C_6$ rotation symmetry.  
Moreover, the $\mac{H}$-$\infty$ lattice 
does not have the full $D_{3d}$ point group symmetry of the 2D honeycomb lattice, because the vector relating adjacent layers, ${\bf a}_1$, is not normal to the honeycomb plane ($a_2 \times c$ or $n_1 \times n_2$ (see Fig. \ref{fig:Harmonic_Lattices})).
This provides additional motivation for studying anisotropic couplings in these systems, even though in the limit that there is no interaction between the layers we expect that the full $D_{3d}$ symmetry will be restored. \cite{Sandilands}
A more detailed discussion of the lattice symmetries is given in Appendix \ref{lattice symmetries}.

	\subsection{The spin liquid ground state}
	
	For a given lattice, the ground state of the Hamiltonian (\ref{Hm}) has the flux pattern that minimizes the energy of the dispersing \mssns. 
 As finding the configuration of fluxes is lattice specific,  here we will separately discuss the solution for each case.  
	
	In the 2D honeycomb lattice (\Hs{\infty}) the ground state has zero flux.\cite{Kitaev} This can be proven exactly using a theorem based on reflection positivity generally referred to as Lieb's theorem. The theorem states that at least one ground state has flux $\Phi_P = (n+2)\pi/4$ through every $n$-sided plaquette that is symmetrically cut by a mirror plane that does not intersect any lattice points. It was originally proven for complex fermions in Ref. \onlinecite{Lieb}, see also Ref. \onlinecite{Macris}. The extension to Majorana fermions was written down recently in Refs. \onlinecite{Chesi} and \onlinecite{Jaffe}. (In contrast to Ref. \onlinecite{Chesi} we refer to the flux as the phase accumulated when hopping around a loop/plaquette following Ref. \onlinecite{Kitaev}.)
	
	As explained in Appendix \ref{sec:lattice}, the 3D \hh lattices' lack  of  mirror planes makes it difficult to uniquely fix the ground state flux configuration.  Numerical searches \cite{Schaffer,Mandal} diagonalizing the \ms Hamiltonian in the thermodynamic limit for flux arrangements consistent with eight-fold enlarged unit cells\cite{Schaffer} or with 216 unit cells and periodic boundary conditions\cite{Mandal} suggest that the ground state on the \Hs{0} lattice lies in zero-flux sector. However, on the \Hs{1} lattice a state with a simple, but non-zero flux pattern was found to have the lowest energy when $J^x = J^y = J^z$. \cite{Schaffer} We follow Ref. \onlinecite{Schaffer}  and call this the $\pi$-flux state.  We have confirmed for the infinite lattice that this state is lower in energy near the isotropic point $J^x = J^y = J^z$ and that the zero-flux state becomes the ground state for small enough $J^x/J^z$ (we keep $J^y = J^x$ throughout) but that the energy difference is very small (see Appendix \ref{sec:lattice} for the ground state energies). 
	
	Once the flux pattern is fixed, the ground state is described by the band structure of the dispersing \mss in this flux background. Certain details of the band structure (though not the ground-state energy or other measurable quantities) are gauge-dependent, so to fix the band structure we work in a particular gauge.
	For the zero flux sector we choose the gauge $u_{\left<ij\right>^\alpha}= i b_i^\alpha b_j^\alpha = 1$ for $i$ on the odd sublattice and $j$ even. If one represents the direction of positive $u_{\left<ij\right>^\alpha}$ by an arrow, then the arrows point from the odd-sites to even-sites.\cite{Hermanns} The other flux sectors are related by changing the sign of the \ms hopping parameter $u_{\left<ij\right>^\alpha}$ on particular links. The $\pi$ flux state on the \Hs{1} lattice can be obtained from this choice by changing the sign of $u_{\left<ij\right>^\alpha}$ on the bonds circled in Fig. \ref{fig:Harmonic_Lattices}.
	 
	  After finding the minimum energy flux sector and fixing an appropriate gauge, the Hamitonian (4) is reduced to a band Hamiltonian for the dispersing $c_j$ \mssns.  In the next subsection we discuss the important features of the resulting band structures, which can be probed directly with Raman scattering.

{\subsection{The fixed-flux Hamiltonian}}

We will now detail the fixed-fluxed eigenstates in terms of the band structures of the \mssns. The Fourier basis  is defined by
 $\mathbf{c}_{\mathbf{k}} = \sum_s e^{i \mathbf{r}_s \cdot \mathbf{k}} \mathbf{c}_{s}$, where    $\mathbf{r}_s$ defines the position of the unit cell $s$,
$\mathbf{c}_{s}=\left(c_{1s},\,c_{3s},...;\,c_{2s},\,c_{4s}...\right)$ defines the vector of distinct \mss in this unit cell.  Note that $\mathbf{c}_ {\mathbf{k}}^\dag = \mathbf{c}_{-{\mathbf{k}}}$,  and ${\mathbf{k}}$ belongs to the first Brillouin  zone (BZ).
    
In the presence of time-reversal symmetry, the \ms Hamiltonian on a bipartite lattice can be written in block off-diagonal form. \cite{Schaffer,Hermanns} We number the sites by their position in the positive $\mathbf{c}-$direction with the odd-numbered sites (yellow) in the first block and the even numbered sites (white) in the second. Then
	\begin{align}
	\mathrm{H}^\eta = \sum_{\textrm{all } {\mathbf{k}}}\mathbf{c}_{-{\mathbf{k}}}^{\mathsmaller T} H_{{\mathbf{k}}}^{\eta,c} \mathbf{c}_{\mathbf{k}} &\hspace{1cm} H_{\mathbf{k}}= \frac{i}{2} \left( \begin{array}{cc}
	0 & \mathcal{D}_{{\mathbf{k}}}^\eta \\
	-\left(\mathcal{D}_{\mathbf{k}}^\eta\right)^\dagger & 0 \\
	\end{array} \right) \nonumber\\
	 \mathbf{c}_{\mathbf{k}}^{\mathsmaller T} &= (c_{1,{\mathbf{k}}},c_{3,{\mathbf{k}}},...;c_{2,{\mathbf{k}}},c_{4,{\mathbf{k}}},...), 
	  \label{Hc}
	\end{align}
		where $\eta$ is a composite index $\eta = n(\Phi)$ for the $n^{\textrm{th}}$-\hh in the flux state $\Phi=\{ \Phi_P\}_P$. If there are 2M sites in the unit cell, $D_{\mathbf{k}}^\eta$ are $M \times M$ matrices. Their precise form for the cases we consider are given in Appendix \ref{Hopping}.
	
	To diagonalize the resulting band Hamiltonian, we express the \mss in terms of complex fermion operators as
		\begin{align} \label{eq:c2a}
	\mathbf{c}_{{\mathbf{k}}} &= \sqrt{2} Q_{\mathbf{k}} \boldsymbol{\alpha}_\mathbf{k} \nonumber \\
	\boldsymbol{\alpha}_{\mathbf{k}}^{\mathsmaller T} &= (a_{1,{\mathbf{k}}},a_{2,{\mathbf{k}}},...;a_{1,-{\mathbf{k}}}^\dagger,a_{2,-{\mathbf{k}}}^\dagger,...) = (\mathbf{a}^{\mathsmaller T}_{\mathbf{k}};\mathbf{a}^\dagger_{-{\mathbf{k}}}),
	\end{align}
where the $a_{n,{\mathbf{k}}}$ satisfy the usual complex fermion algebra $\{a^\dag_{n,{\mathbf{k}}},a_{m,{\bf q}}\} = \delta_{nm} \delta_{{\mathbf{k}}{\bf q}}$ and $Q_{\mathbf{k}}$ is a unitary matrix that diagonalizes the matrix $H^\eta_{\mathbf{k}}$:
	\begin{align}
	\mathrm{H}^\eta = \sum_{\textrm{all } {\mathbf{k}}}\boldsymbol{\alpha}_{-{\mathbf{k}}}^{\mathsmaller T} \tilde{H}_{{\mathbf{k}}}^{\eta} \boldsymbol{\alpha}_{\mathbf{k}}  
	\hspace{1 cm} \tilde{H}_{{\mathbf{k}}}^{\eta} = 2 Q_{\mathbf{k}}^\dagger H_{\mathbf{k}}^\eta Q_{\mathbf{k}}.
	\end{align}	
	The diagonalized Hamiltonian reads as
	\begin{align}\label{Ha}
	\mathrm{H}^\eta = \frac{1}{2}\sum_{m=1,...,M ; \mathbf{k}} \varepsilon_{m,{\mathbf{k}}} (a_{m,{\mathbf{k}}}^\dagger a_{m,\mathbf{k}} - a_{m,{\mathbf{k}}}a_{m,{\mathbf{k}}}^\dagger ),
	\end{align}
	where $m$ labels the bands. 
	
	The fact that $a^\dagger_{n,{\mathbf{k}}}$ is related to $a_{n,{\mathbf{k}}}$ by hermitian conjugation leads to some redundancy. This, along with the condition $c^\dagger_{\mathbf{k}} = c_{-{\mathbf{k}}}$, requires that $Q^{\mathsmaller T}_\mathbf{k} = \gamma Q^\dagger_{-{\mathbf{k}}}$, where $\gamma = \left(
	\begin{array}{cc}
	0 & \1 \\
	\1 & 0 \\
	\end{array}
	\right)$.
	In practice we impose this constraint by computing $Q_{\mathbf{k}}$ for ${\mathbf{k}}$ in half of the Brillouin zone and defining $Q_{-{\mathbf{k}}} = \gamma Q^*_{\mathbf{k}}$. 
	The ground state is the one with no quasi-particles, $a_{m,{\mathbf{k}}} \ket{0} = 0$, and has energy $E_0 = -\frac{1}{2}\sum_{m;{\mathbf{k}}} \varepsilon_{m,{\mathbf{k}}}$. 
	
	For every two \mss there is one spinless fermion, whose corresponding excited state $a_{m,-{\mathbf{k}}}^\dagger \ket{0}$ has energy $\varepsilon_{m,{\mathbf{k}}}$. We refer to the energies of the different quasiparticles as different \ms bands, of which there are one for every two sites in the unit cell. 
The physical excitations above the ground state correspond to pairs of the original Majoranas: flux+\ms or two \mssns.  Operators that create single-Majorana fermion excitations do not commute with the constraint, and therefore are not physical.\cite{GSfoot} 
	
The band structures that result from this picture have some striking similarities for all three of the lattices considered here. First, there are two distinct spin liquid phases: a gapless phase centered at the isotropic point $J^x =J^y=J^z$, and a gapped phase when $J^z> (J^x + J^y)$. Second, the codimension of the Fermi surface in the gapless phase is always 2:\cite{infinite-D,Lee,Hermanns} it is a Dirac line in 3D, and a Dirac point on the 2D honeycomb lattice.  As a result, in the gapless phase, the low-energy density of states asymptotically obeys the power law $\rho(\omega) \sim \omega$.  
\\
	
	\section{Raman response} \label{RamanSec}

 In this section we present the derivation of the Raman response and the polarization dependence of the Raman spectrum for the Kitaev spin liquid state on three \hh lattices: the \Hs{\infty}, \Hs{0}, and \Hs{1} corresponding to $\alpha$-, $\beta$-, and $\gamma$-Li$_2$IrO$_3$, respectively. While our study does not represent a complete microscopic theory of Raman spectra (including all possible contributions), it should provide experimentalists with the main signatures of magnetic quantum number fractionalization that might be visible in these compounds. 
		
	Raman scattering is the inelastic scattering of light by energies in the meV range. It measures correlations between two-photon events (one ingoing and one outgoing). By Fermi's golden rule the Raman intensity can be written as the Fourier transform of a correlation function
		\begin{align}\label{I1}
		I(\omega) &= \int dt e^{i \omega t} \braket{{R}(t){R}(0)},
		\end{align}
		where $R(t) = e^{iHt}\,R\,e^{-iHt}$ is the Raman operator in the Heisenberg representation. 
We derive the Raman  operator using the Loudon and Fleury (LF) approach for Mott insulators,\cite{Fleury} which describes the effective interaction of light with spin degrees of freedom. The dominant coupling was shown to be the electric one.\cite{Fleury} In these systems, the effective interaction can be obtained by performing a hopping expansion; the leading term is the LF Hamiltonian.\cite{Ko}

If the incoming photon frequency is smaller than the appropriate Mott gap,\cite{Shastry2} the resulting LF Hamiltonian turns out to be precisely the exchange Hamiltonian augmented with polarization-dependent terms corresponding to the component of each photon's electric field along the bond that the electron virtually hops across. For a spin-exchange Hamiltonian $H$ this leads to\cite{Perkins,Ko}
		\begin{align}
		H & = \sum_{i,j;\alpha,\beta} \Gamma_{ij}^{\alpha \beta} \sigma^\alpha_i \sigma^\beta_j \label{Hex}\\
		{R} &= \sum_{i,j;\alpha,\beta} (\boldsymbol{\epsilon}_{\textrm{in}} \cdot \mathbf{d}_{ij}) (\boldsymbol{\epsilon}_{\textrm{in}} \cdot \mathbf{d}_{ij}) \Gamma_{ij}^{\alpha \beta} \sigma^\alpha_i \sigma^\beta_j. \label{R}
		\end{align}
		where $\boldsymbol{\epsilon}_{\textrm{in}}$ and $\boldsymbol{\epsilon}_{\textrm{out}}$ are the polarization vectors of the incoming and outgoing light,  $\mathbf{d}_{ij}$ is the vector from site $i$ to site $j$, $\sigma_i^\alpha$ is the $\alpha^{th}$ component of the spin on site $i$, and $\Gamma$ is a generalized exchange constant.

	\subsection{Quasiparticle picture}
	As emphasized in the introduction, one striking feature of Raman spectroscopy in a Kitaev spin liquid is that the Raman operator couples only to the dispersing \mssns.
	For the Kitaev Hamiltoian (\ref{H}), the LF operator Eq.~(\ref{R}) takes the form \cite{Perkins}
	\begin{align}\label{R2}
	R &= \sum_{\left<ij\right>^\alpha} (\boldsymbol{\epsilon}_{\textrm{in}} \cdot \mathbf{d}^\alpha) (\boldsymbol{\epsilon}_{\textrm{out}} \cdot \mathbf{d}^\alpha)
	J^\alpha \sigma_i^\alpha \sigma_j^\alpha \nonumber\\& = i\sum_{\left<ij\right>^\alpha} (\boldsymbol{\epsilon}_{\textrm{in}} \cdot \mathbf{d}^\alpha) (\boldsymbol{\epsilon}_{\textrm{out}} \cdot \mathbf{d}^\alpha) J^\alpha 
	u_{\left<ij\right>^\alpha}  c_i c_j,
	\end{align}
	which is a simple quadratic operator in terms of the \mss $\mathbf{c}_\mathbf{k}$. 	
	Remarkably, due to its similarity to the original Kitaev spin Hamiltonian, the Raman operator (\ref{R2}) does not excite the gapped flux excitations. 	This is one distinct advantage of using Raman response in these systems, as it probes only one of the fractionalized sectors.  (In contrast, neutron scattering always excites both fluxes and \mssns.)  In addition, the fact that the Raman operator conserves the flux in each plaquette greatly simplifies the calculations of the Raman intensity, since we can use the fixed-flux Hamiltonian (\ref{Hc}).
	Consequently, we can write the Raman operator in terms of the Bogoliubov-deGennes fermions that diagonalize the Hamiltonian (see Eq.~(\ref{eq:c2a})). 
	In this basis, the Raman operator takes the form $R = \sum_\mathbf{k} R_\mathbf{k}$ with
	\begin{align}\label{Rk}
	R_\mathbf{k} = A_{mn,\mathbf{k}} a^\dagger_{m,\mathbf{k}}a_{n,\mathbf{k}} + \frac{1}{2}\left(B_{mn,\mathbf{k}} a^\dagger_{m,\mathbf{k}} a^\dagger_{n,-\mathbf{k}} +  h.c.\right),
	\end{align}
	where indices $m,n=1,...,M$ refer to bands and $A_{mn,\mathbf{k}}$ and $B_{mn,\mathbf{k}}$ are bilinear functions of the in and out polarizations. 
	One can reduce the Raman intensity to a weighted   two-\ms DOS (2-DOS) 	given by
	\begin{align} \label{2PartDOS}
	\rho_2(\omega) &= \sum_{m,n;\mathbf{k}} \delta(\omega-\varepsilon_{m,\mathbf{k}}-\varepsilon_{n,\mathbf{k}}).
	\end{align}
	Specifically, we obtain
	\begin{align} \label{result}
	I(\omega) &= \pi \sum_{m,n;\mathbf{k}}  \delta(\omega - \varepsilon_{m,\mathbf{k}} - \varepsilon_{n,\mathbf{k}}) |B_{mn,\mathbf{k}}|^2
	\end{align}
	Note that $B_{mn,\mathbf{k}}$ is the antisymmetric matrix corresponding to the creation of two excitations in Eq.~(\ref{Rk}). The $A_{mn,\mathbf{k}}$ part annihilates the ground state and thus does not contribute to the Raman response. 
		
	Eq.~(\ref{result}) describes the response at zero temperature.  At finite temperature the $A_{mm',\mathbf{k}}$ can contribute to the correlation function by relaxing thermally excited quasiparticles. In general the finite temperature dependence is complicated by the presence of thermally excited fluxes, which are strongly coupled to the \mssns.  However, for temperatures well below the flux gap one can restrict the calculation to the zero flux sector.  In this case Eq.~(\ref{result}) generalizes to
	\begin{align}
	& I(\omega) = 2\pi \sum_{m,n;\mathbf{k}} \left\{A_{mn,\mathbf{k}} A_{nm,\mathbf{k}} \delta\left(\omega + \left[\varepsilon_{m,\mathbf{k}}-\varepsilon_{n,\mathbf{k}}\right]\right) \phantom{\frac{1}{2}} \right. \nonumber\\
	& \hspace{3cm} \times n_F(\varepsilon_{m,\mathbf{k}})n_F(-\varepsilon_{n,\mathbf{k}}) \nonumber 
	\\ 
	&\left. + \frac{1}{2}|B_{mn,\mathbf{k}}|^2 \delta\left(\omega - \left[\varepsilon_{m,\mathbf{k}}+\varepsilon_{n,\mathbf{k}}\right]\right) n_F(-\varepsilon_{m,\mathbf{k}})n_F(-\varepsilon_{n,\mathbf{k}}) \right\},
	\end{align}
	where 
	$n_F(\varepsilon) = \frac{1}{1+e^{\beta \varepsilon}}$ is the Fermi-Dirac distribution.

	\subsection{Polarization dependence} \label{PolSec}
	
	In the following we will identify the distinctive features of the polarization dependence of the magnetic excitations in the Kitaev model and their relation to the lattice structure and the form of the electronic Hamiltonian. But first we 
	review the arguments for the polarization dependence that follow directly from lattice symmetry and the form of the LF operator. Due to the generality of these assumptions, the relationships obtained are expected to hold for any nearest-neighbor spin-exchange Hamiltonian on these lattices, such as the Heisenberg-Kitaev model. Ways to break these constraints on polarization dependence are discussed in the next section.
	
	The form of the Raman operator $R$ depends on the polarization of incoming and outgoing light. For the LF operator [Eq.~(\ref{R})] this is because driving an exchange process along a certain bond requires the photon's polarization to have a component along that bond vector. 
	We will use the short hand notation $\mu \nu$ to refer to a scattering geometry with the polarization of incoming light along $\hat{\mathbf{\mu}}$ and outgoing along $\hat{\mathbf{\nu}}$. 
	It is convenient to use the cubic coordinates $\mu,\nu = a,b,c$ (see Fig. \ref{fig:Harmonic_Lattices}) due to their relation to the symmetries of the lattice (see Appendix \ref{lattice symmetries}).

	Since the Raman operator is linear in polarizations it can always be written as a tensor dotted with the polarization vectors.   
	\begin{align} \label{Rp}
	R 
	&=\sum_{\mu,\nu= a,b,c}  (\boldsymbol{\epsilon}_{\textrm{in}})_\mu R_{\mu \nu} (\boldsymbol{\epsilon}_{\textrm{out}})_\nu\, . 
	\end{align}
	For a spin-exchange Hamiltonian the Raman operators are
	\begin{align}\label{Rp2}
	R_{\mu \nu} &\equiv  \sum_{\alpha=\text{bonds}} d_{\mu}^\alpha H^\alpha d_{\nu}^\alpha,
\end{align}
	where $\alpha$ refers to a particular bond direction, $d^\alpha_\mu$ is the $\mu^{\text{th}}$ component of the bond vector $d^\alpha$, and $H^\alpha$ is the sum of the spin-exchange terms on bonds directed along $\alpha$. 
	
	By inserting Eq.~(\ref{Rp}) into Eq.~(\ref{I1}) the Raman intensity can be decomposed into a linear combination of intensities obtained for pairs of these Raman operators. 		
	\begin{align}
	I(\omega) &= \int dt e^{i \omega t} \braket{ R(t) R(0)} \nonumber \\
	& \equiv \sum_{\mu, \nu, \mu', \nu'}  (\boldsymbol{\epsilon}_{\textrm{in}})_\mu (\boldsymbol{\epsilon}_{\textrm{out}})_\nu (\boldsymbol{\epsilon}_{\textrm{in}})_{\mu'} (\boldsymbol{\epsilon}_{\textrm{out}})_{\nu'} I_{\mu\nu,\mu'\nu'}(\omega),
	\end{align}
	where
	\begin{align}\label{I-terms}
	I_{\mu\nu,\mu'\nu'}(\omega) &\equiv \int dt e^{i \omega t} \frac{1}{2}\left( \braket{R_{\mu\nu}(t) R_{\mu'\nu'}(0) }  \right. \nonumber\\ & \left.  
	\hspace{2cm} + \braket{R_{\mu'\nu'}(t) R_{\mu\nu}(0) } \right)
	\end{align}
	The terms with $\mu,\nu\ne \mu',\nu'$ only appear when either the incoming or outgoing polarization is not along a cubic unit vector $\hat{a}$, $\hat{b}$, or $\hat{c}$ so that the Raman operator is the sum of multiple terms. For instance, a polarization with $\boldsymbol{\epsilon}_{\textrm{in}} = \hat{\mathbf{a}}$ and $\boldsymbol{\epsilon}_{\textrm{out}} = \frac{1}{\sqrt{2}}(\hat{\mathbf{a}} + \hat{\mathbf{b}})$ would give
	\begin{align}\label{polex}
	I = \frac{1}{2} \left(I_{aa,aa} + I_{ab,ab}\right) + \sqrt{2} I_{aa,ab},
	\end{align}
	where we have left the $\omega$ dependence implicit, to simplify the expression. One could measure the spectrum $I_{aa,ab}$ by first isolating the other terms $I_{aa,aa}$ and $I_{ab,ab}$ and then subtracting off the components of these other two from the spectrum measured with the polarization in Eq.~(\ref{polex}).

Note that the intensity ``cross terms" $I_{ab,cd}, ab \ne cd$ are sums and differences of measurable quantities and may therefore be positive or negative. 
	Otherwise, we refer to the intensity for a simple polarization configuration along the cubic directions $\mu$ and $\nu$ by $I_{\mu\nu}$, which is given by $I_{\mu\nu} \equiv I_{\mu\nu,\mu\nu}$ above. 
	
	The space group symmetries reduce the number of independent spectra, which can be classified by the irreducible representation (irrep) under which the point group acts on the Raman tensor $R_{\mu\nu}$. The cross terms between different irreps are zero since symmetry must act trivially on the spectrum $I$. As discussed in more detail in Appendix \ref{pol}, the point group $D_{2h}$ reduces the number of non-zero spectra of the 3D lattices to nine (because 12 cross terms vanish): 
	\begin{align}\label{nonzero}
	I_{aa}, I_{bb}, I_{cc}, I_{aa,bb}, I_{aa,cc}, I_{bb,cc}, I_{ab}, I_{ac}, I_{bc} \ne 0.
	\end{align}
	The first three correspond to different representations of the $A_{1g}$ irrep, the next three are their cross terms, and the remaining three correspond to the $B_{1g}$, $B_{2g}$ and $B_{3g}$ irreps respectively. 
	 
	In addition to the point group symmetries, the 3D \hh lattices have an additional constraint that can be understood in terms of a non-symmorphic screw symmetry of a related lattice of the same connectivity. This reduces the number of independent spectra to 6 by the relations: 
	\begin{align}\label{screw}
	4I_{aa} = I_{bb}, \hspace{.7cm} 2I_{aa,cc} = I_{bb,cc}, \hspace{.7cm} 2I_{ac} = I_{bc}.
	\end{align}
	We discuss this in detail in the results section below. 
		
	Quite separately, the close resemblance of the LF operator Eq.~(\ref{R}) to the spin-exchange Hamiltonian can give additional linear relations between the spectra. This is due to the fact that the Hamiltonian acts trivially on eigenstates. Therefore if two Raman operators sum to the Hamiltonian, their spectra must be related. We refer to this as the Loudon-Fleury (LF) relationship. For the remaining six spectra, this type of argument leads to
	\begin{align}\label{LF3D}
	I_{cc} = 9 I_{aa} = - 3 I_{aa,cc}
	\end{align} 
	for a nearest-neighbor spin-exchange Hamiltonian on the 3D lattices. That leaves only three independent spectra for the model considered here. We choose to plot $I_{aa}$, $I_{ab}$, and $I_{ac}$, which represent the $A_{1g}$, $B_{1g}$, and $B_{2g}$ channels. Then the other channels are either zero (see Eq.~\ref{nonzero}), or are linearly related to these ones by Eqs.~(\ref{screw}-\ref{LF3D}). In particular, $B_{3g}$ is degenerate with $B_{2g}$. 
	
	For the 2D lattice with full lattice symmetry there are two independent spectra by point group symmetry and one LF-relation ($R_{A_{1g}} = H$). This reduces it to one independent spectrum -- the $E_g$ channel, which we represent with $I_{xx}$. \cite{Perkins} In Appendix \ref{pol} we review this argument as well as treating the case of bond anisotropy $J^z \ne J^x = J^y$ which breaks the $C_6$ rotation symmetry. We find that there are then four independent spectra by symmetry and the same LF-relation reduces this to two.  
	In that case we plot $I_{xx}$ and $I_{xy}$ which represent the $A_{1g}$ and the $B_{1g}$ channels respectively, of $D_{2h}$. 
	Inclusion of next-nearest-neighbor spin-exchange couplings does not change the vanishing of the $A_{1g}$ channel due to the LF-relationship, although it would break the two equalities in Eq.~(\ref{LF3D}) for the 3D lattice. See Appendix \ref{pol} for more details.

 	\begin{figure*}[htb] 
\begin{minipage}{0.32\linewidth}
		\begin{tikzpicture}
		\node[anchor=south west,inner sep=0] (image) at (0,0) {\includegraphics[width=58mm, trim = 0mm 0mm 10mm 9mm, clip]{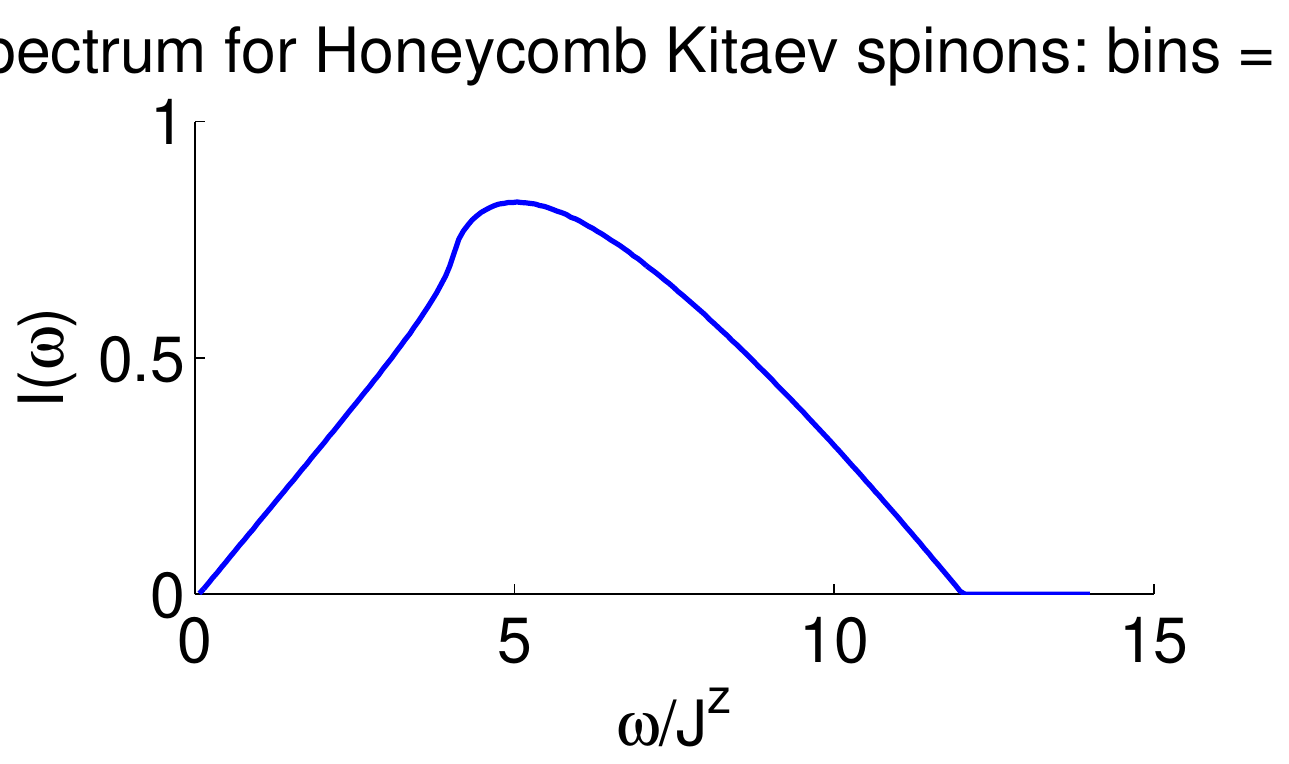}};
		\begin{scope}[xshift=0.7mm,yshift=27.5mm]
		\clip [rounded corners=0mm] (0,0) rectangle coordinate (centerpoint) (32mm,25mm);
		\node [inner sep=0] (image) at (centerpoint) {\includegraphics[width=30mm, trim = 1.8mm -2.0mm 42mm 12mm, clip]{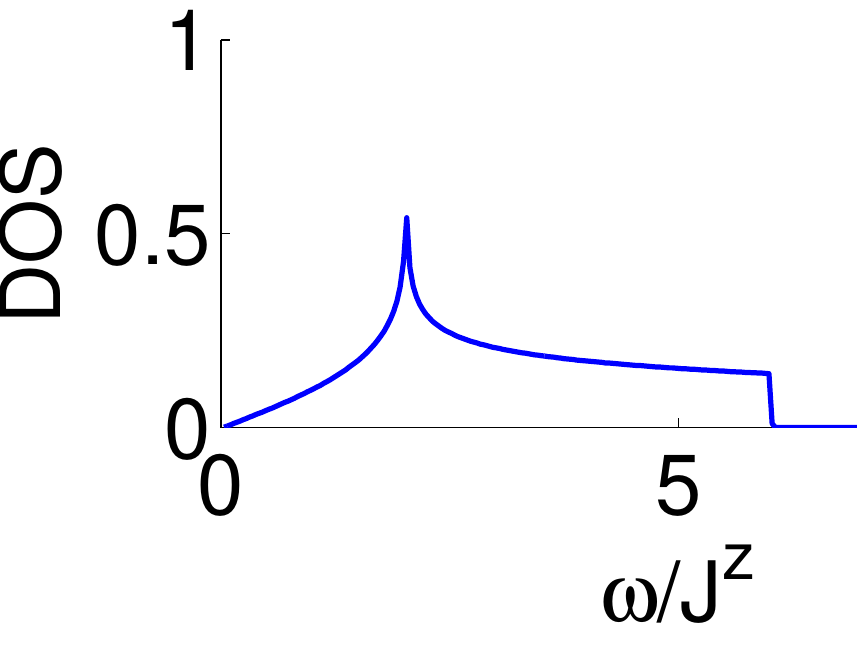}};
		\end{scope}
		\begin{scope}[xshift=28.5mm,yshift=27.8mm]
		\clip [rounded corners=0mm] (0,0) rectangle coordinate (centerpoint) (32mm,25mm);
		\node [inner sep=0] (image) at (centerpoint) {\includegraphics[width=25mm, trim = 1.8mm 1.5mm 35mm 2mm, clip]{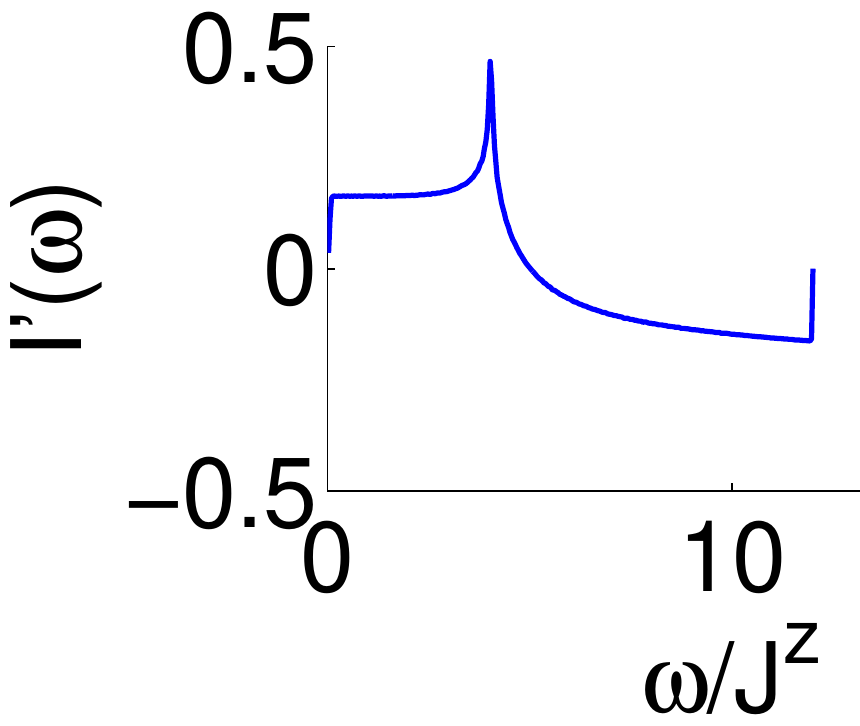}};
		\end{scope}
		\node[anchor=south west,inner sep=0] (image) at (4.7,2.2) {(a)};
		\end{tikzpicture}
\end{minipage}	
\begin{minipage}{0.32\linewidth}
		\begin{tikzpicture}
		\node[anchor=south west,inner sep=0] (image) at (0,0) {\includegraphics[width=56mm, trim = 0mm 0mm 10mm 9mm, clip]{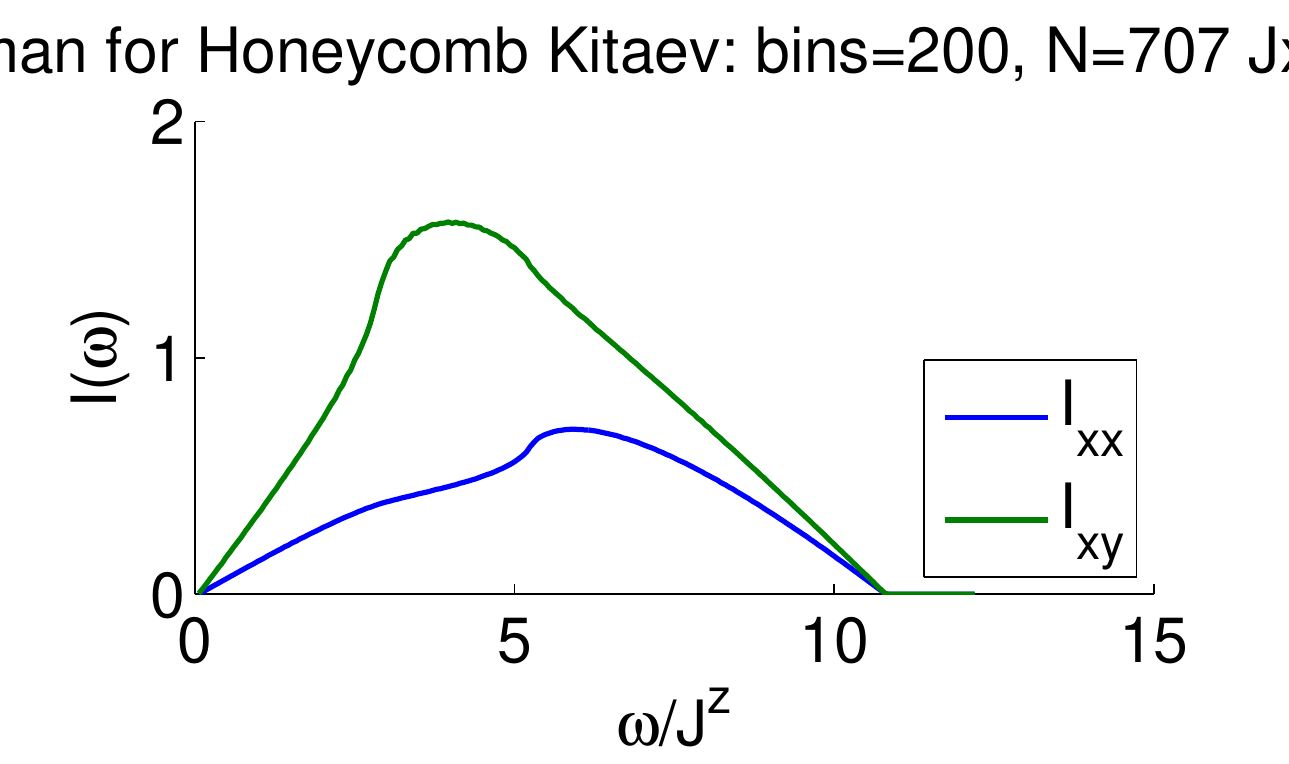}};
		\begin{scope}[xshift=-23.5mm,yshift=5.0mm]
		\node [inner sep=0] (image) at (4,3.5) {\includegraphics[width=30mm, trim = 1.8mm -2.0mm 42mm 12mm, clip]{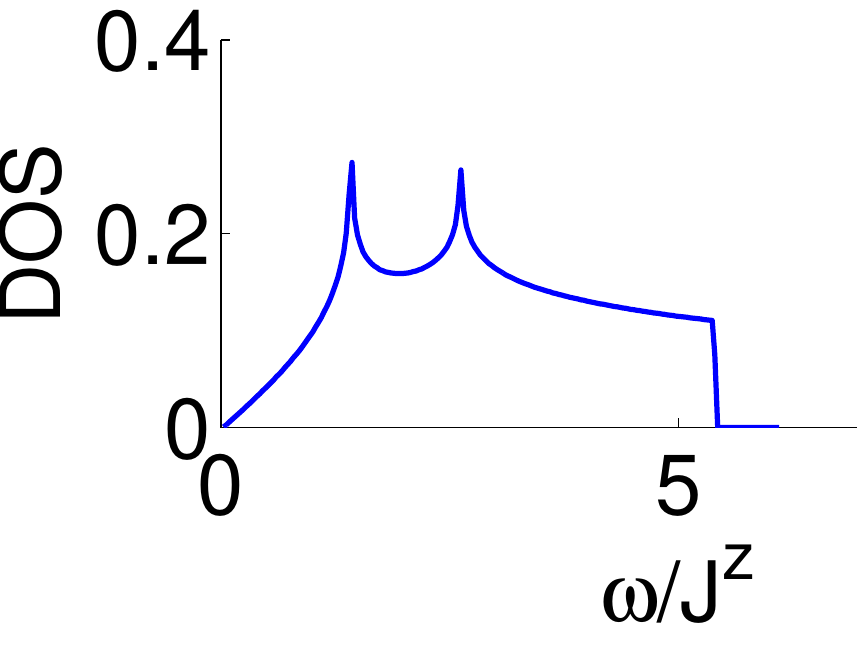}};
		\end{scope}
		\begin{scope}[xshift=28.5mm,yshift=27.8mm]
		\clip [rounded corners=0mm] (0,0) rectangle coordinate (centerpoint) (32mm,25mm);
		\node [inner sep=0] (image) at (centerpoint) {\includegraphics[width=25mm, trim = 1.8mm 1.5mm 35mm 2mm, clip]{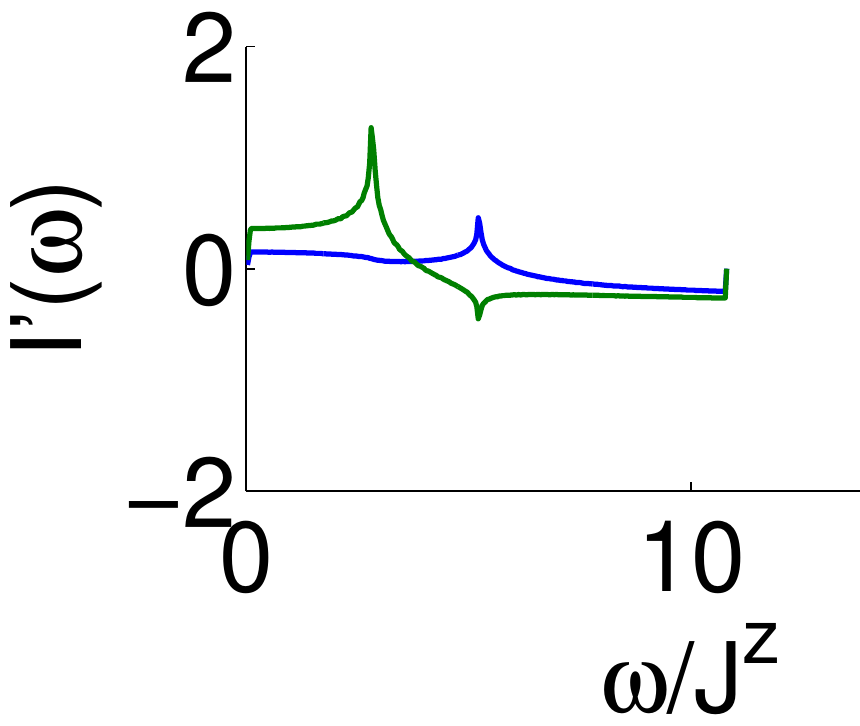}};
		\end{scope}
		\node[anchor=south west,inner sep=0] (image) at (4.7,2.2) {(b)};
		\end{tikzpicture}	 			
\end{minipage}
\begin{minipage}{0.32\linewidth}
		\begin{tikzpicture}
		\node[anchor=south west,inner sep=0] (image) at (0,0) {\includegraphics[width=56mm, trim = 0mm 0mm 15mm 9mm, clip]{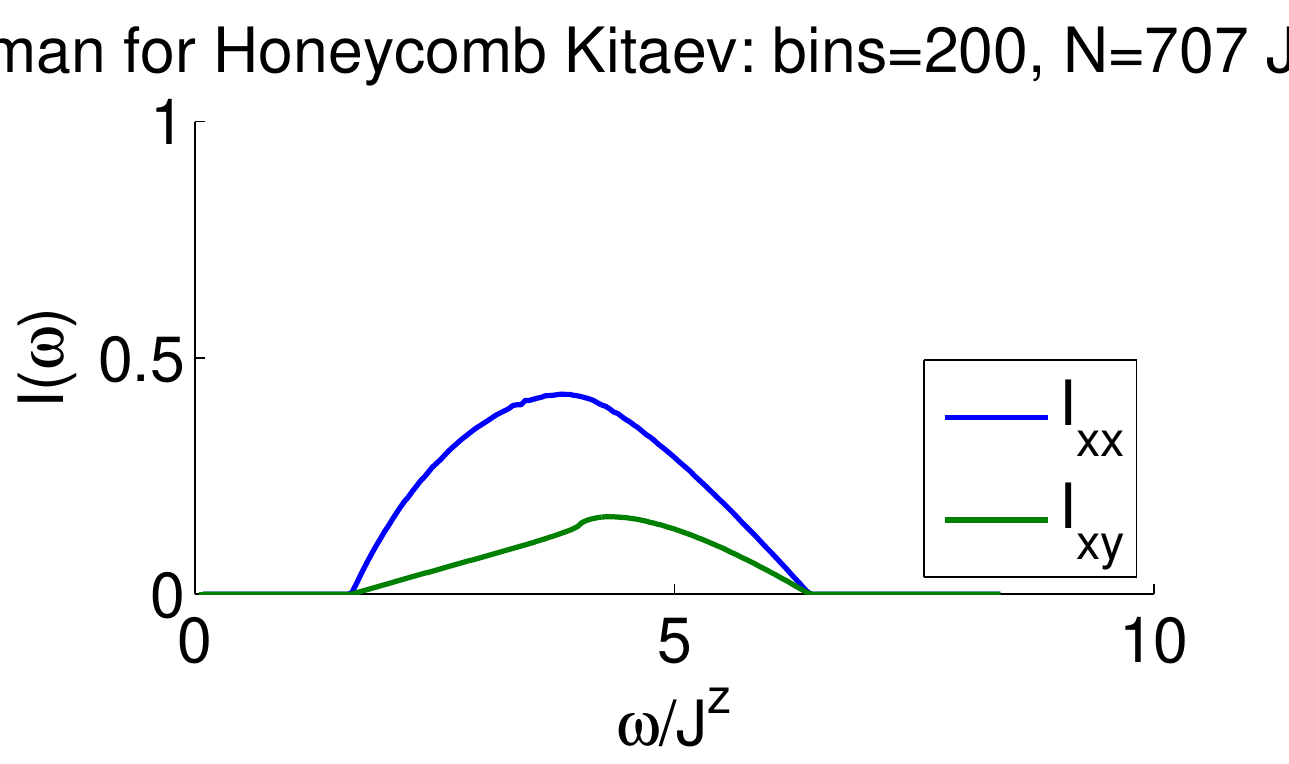}};
		\begin{scope}[xshift=-23.1mm,yshift=5.0mm]
		\node [inner sep=0] (image) at (4,3.5) {\includegraphics[width=30mm, trim = 1.8mm -2.0mm 42mm 12mm, clip]{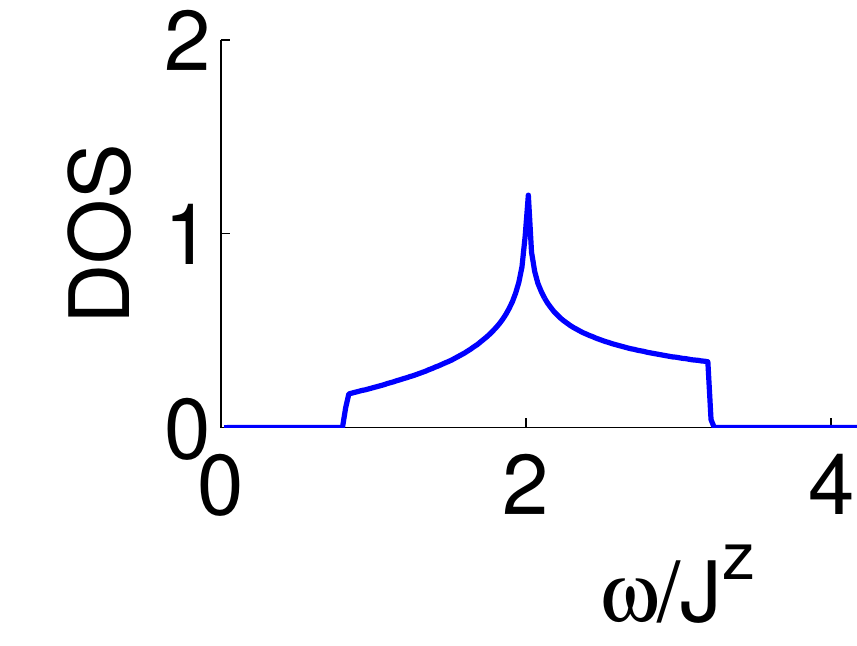}};
		\end{scope}
		\begin{scope}[xshift=28.5mm,yshift=27.8mm]
		\clip [rounded corners=0mm] (0,0) rectangle coordinate (centerpoint) (32mm,25mm);
		\node [inner sep=0] (image) at (centerpoint) {\includegraphics[width=25mm, trim = 1.8mm 1.5mm 35mm 2mm, clip]{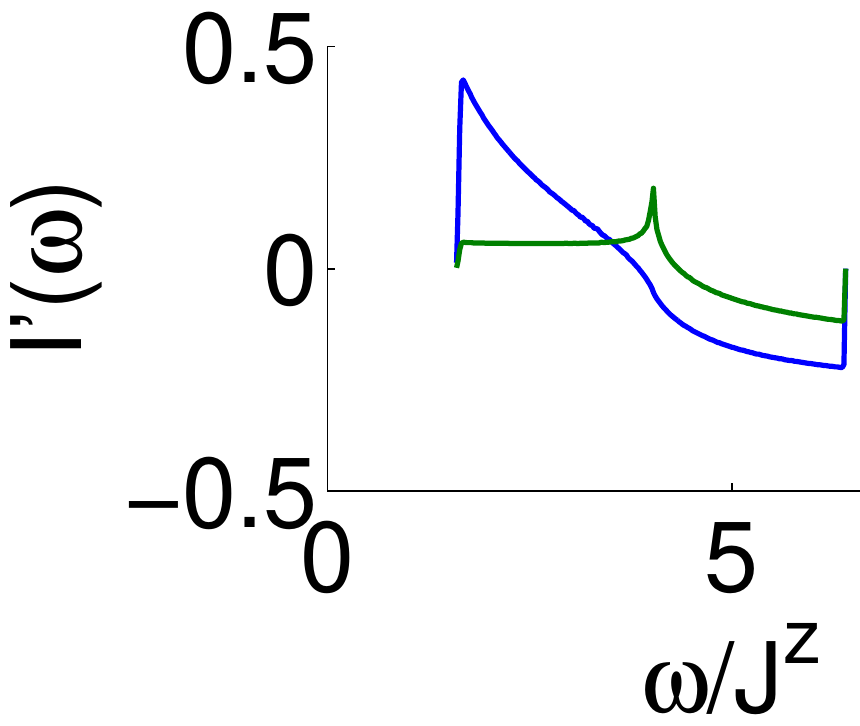}};
		\end{scope}
		\node[anchor=south west,inner sep=0] (image) at (4.9,2.2) {(c)};
		\end{tikzpicture}
\end{minipage}	
 		\caption{(Color online) Raman intensities for the pure Kitaev model on the 2D honeycomb lattice  computed for: (a) the symmetric point \mbox{$J^x = J^y = J^z$}, (b) another gapless point \mbox{$J^x = J^y = 1.43J^z$}, (c) a gapped point \mbox{$J^x = J^y = 0.3J^z$}.  For each figure, the inset on the  left is a  plot of the one-particle DOS;  the inset  on the right  is a  plot of the derivatives of the Raman intensity $I'(\omega)$.  }
 		\label{2D}
 	\end{figure*}
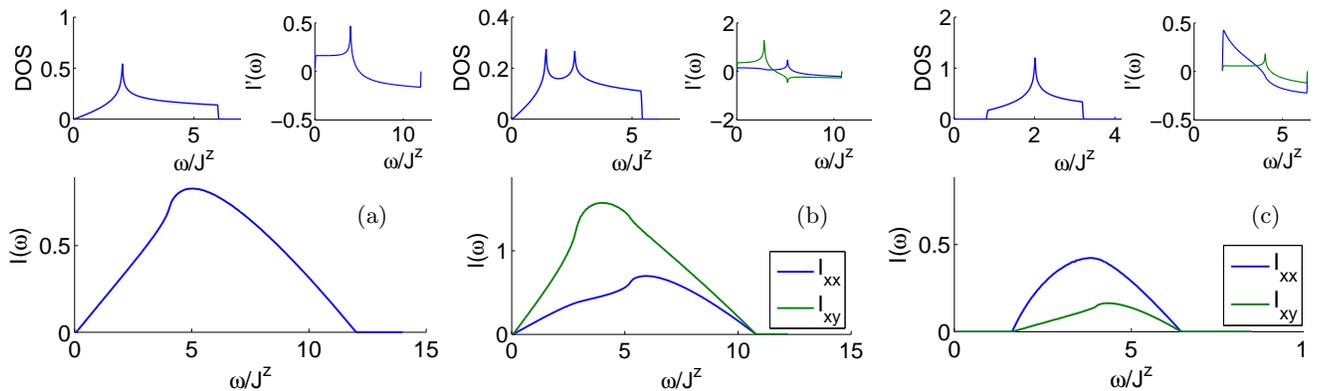

	\section{Results} \label{ResultsSec}
	
	In this section, we present our results and discuss the key features of the Raman response of the Kitaev model on the \Hs{0}, \Hs{1}, and \Hs{\infty} lattices. For all three lattices the Raman response differs qualitatively from the structure factor. This follows from the fact that Raman directly probes the fermion DOS, without coupling to fluxes. The difference is especially pronounced in the gapless phase, where the Raman spectrum is gapless but the structure factor is gapped. In addition, we will highlight certain distinctive features of the polarization dependence in the Raman spectra for a Kitaev spin liquid phase on these lattices.
	
 	\subsection{2D Honeycomb spectra}
 	
 	The simplest case is the 2D honeycomb lattice. It has one \ms band and at most four independent polarization combinations by symmetry.
 	At the isotropic point $J^x = J^y = J^z$ the $C_6$ rotational symmetry allows only two independent Raman spectra corresponding to the channels $A_{1g}$ and $E_g$, which are probed by the combinations
\begin{align}
I_{A_{1g}} &= \frac{1}{2}\left(I_{xx} - I_{xy}\right) \\
I_{E_g} &= I_{xy}.
\end{align} 
As discussed in the previous section, there is no response in the $A_{1g}$ symmetry channel (it does not couple to the spins) due to the LF relationship. For equal couplings, the rotation symmetry gives a polarization independent spectrum corresponding to the $E_{g}$ channel. This is a peculiar prediction that holds for any symmetry-preserving spin-exchange Hamiltonian with a ground state that also preserves the lattice symmetry. In particular it expected to hold in the Kitaev spin-liquid phase in the presence of a Heisenberg exchange perturbation. The resulting spectrum for the pure Kitaev model is plotted in Fig. \ref{2D}(a).

 	\begin{figure}[htb]
 		\centering
 		\includegraphics[width=.9\linewidth, trim = 0mm 0mm 0mm 8mm, clip]{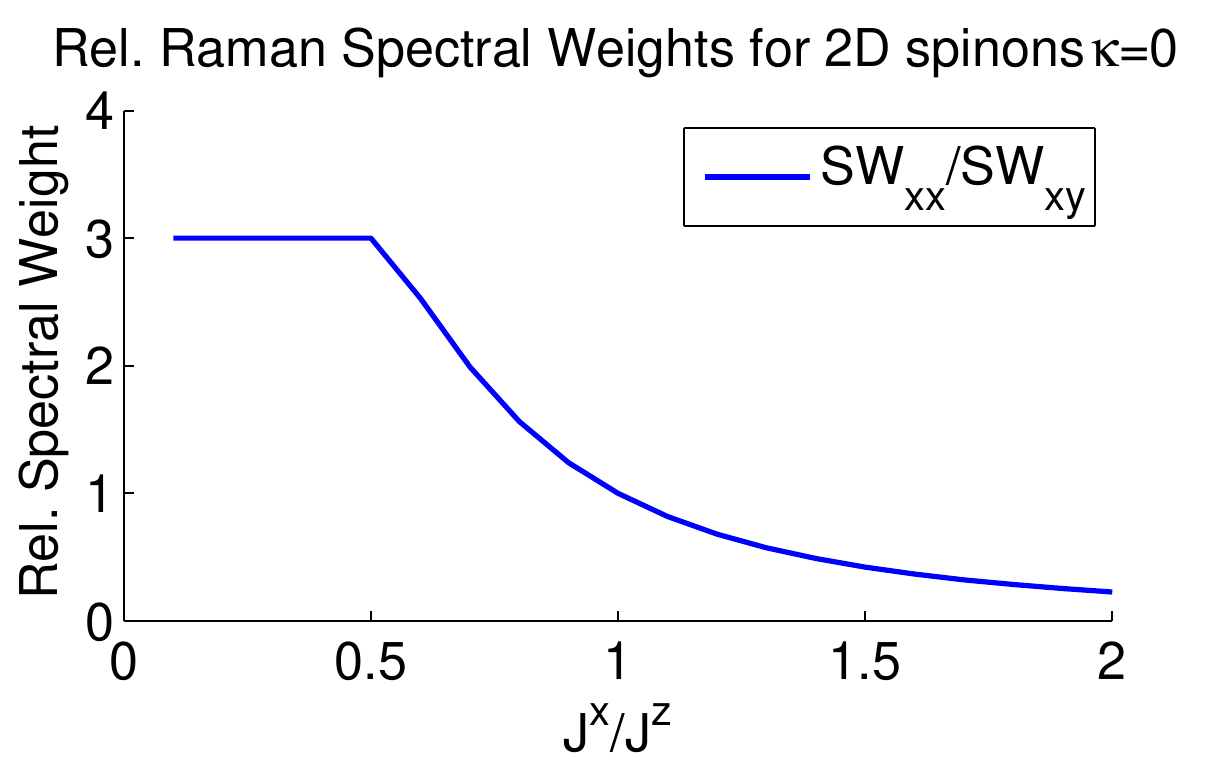}
 		\caption{The relative total spectral weight $SW = \int I(\omega)d\omega$ of the two representative Raman intensities as a function of the Kitaev exchange-coupling anisotropy for the 2D honeycomb lattice. The transition to the gapless phase occurs at \mbox{$J^x/J^z = 0.5$}.}
 		\label{weights}
 	\end{figure}
 		
 		\begin{figure*}[htb]
 			\centering
 			\begin{minipage}{.32\linewidth} 
 				\begin{tikzpicture}
 			\node[anchor=south west,inner sep=0] (image) at (0,0) {
	 			\includegraphics[width=\linewidth, trim = 0mm 0mm 0mm 0mm, clip]{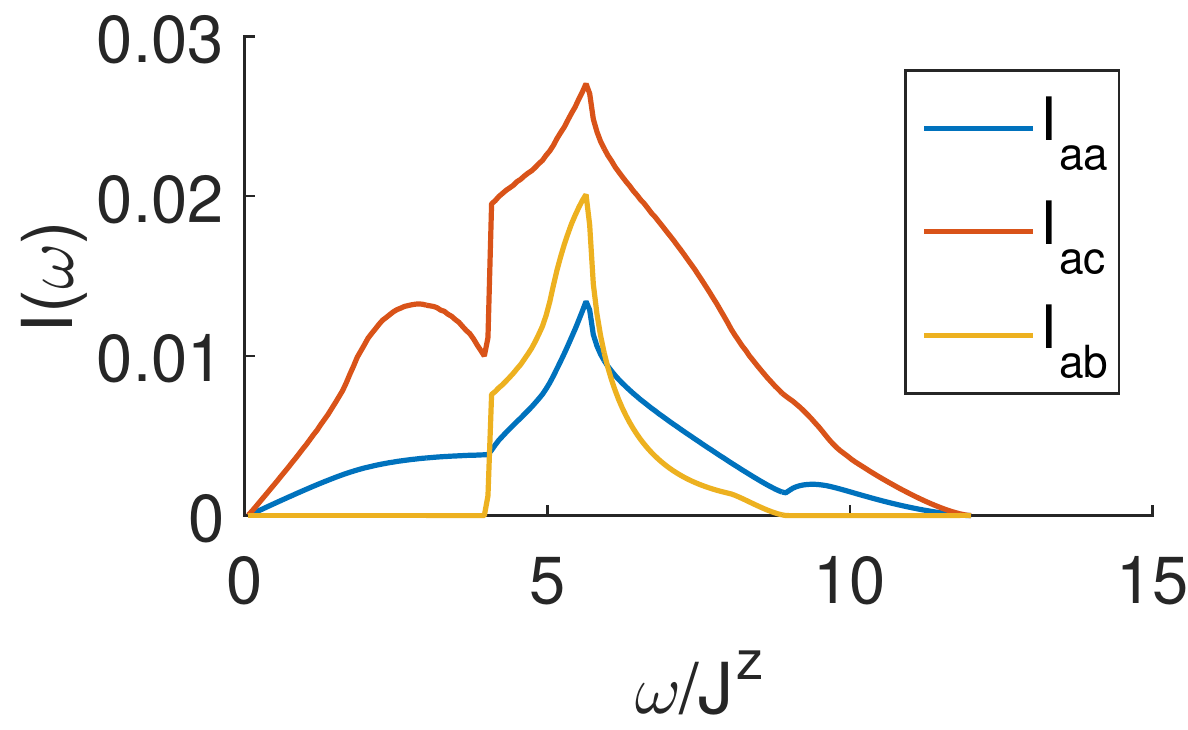} }; 
	 		\node[anchor=south west,inner sep=0] (image) at (1.4,3.17) {(a)};
	 			\end{tikzpicture}
 			\end{minipage} 
 			\begin{minipage}{.32\linewidth}
	 			\begin{tikzpicture}
 			\node[anchor=south west,inner sep=0] (image) at (0,0) {
 				\includegraphics[width=\linewidth, trim = 0mm 0mm 0mm 0mm, clip]{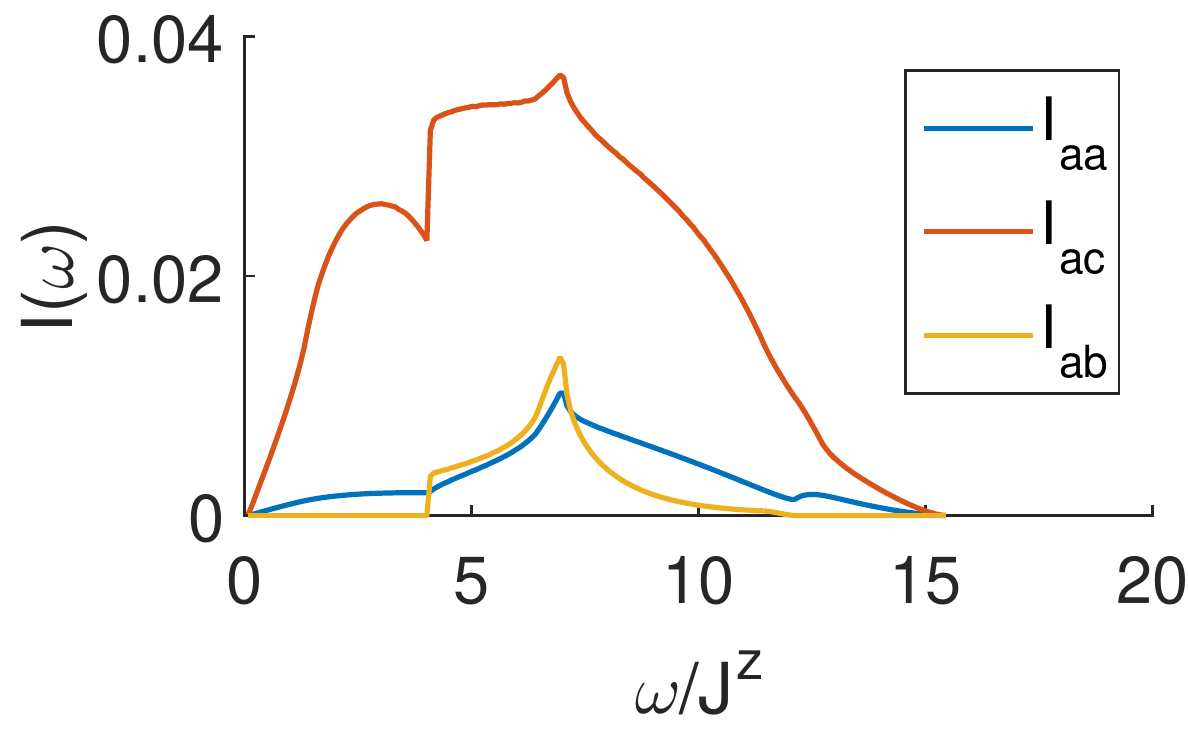} }; 
 			\node[anchor=south west,inner sep=0] (image) at (1.4,3.17) {(b)};
	 			\end{tikzpicture}
	 		\end{minipage}
 			\begin{minipage}{.32\linewidth} 
	 			\begin{tikzpicture}
 			\node[anchor=south west,inner sep=0] (image) at (0,0) {
 				\includegraphics[width=\linewidth, trim = 0mm 0mm 0mm 0mm, clip]{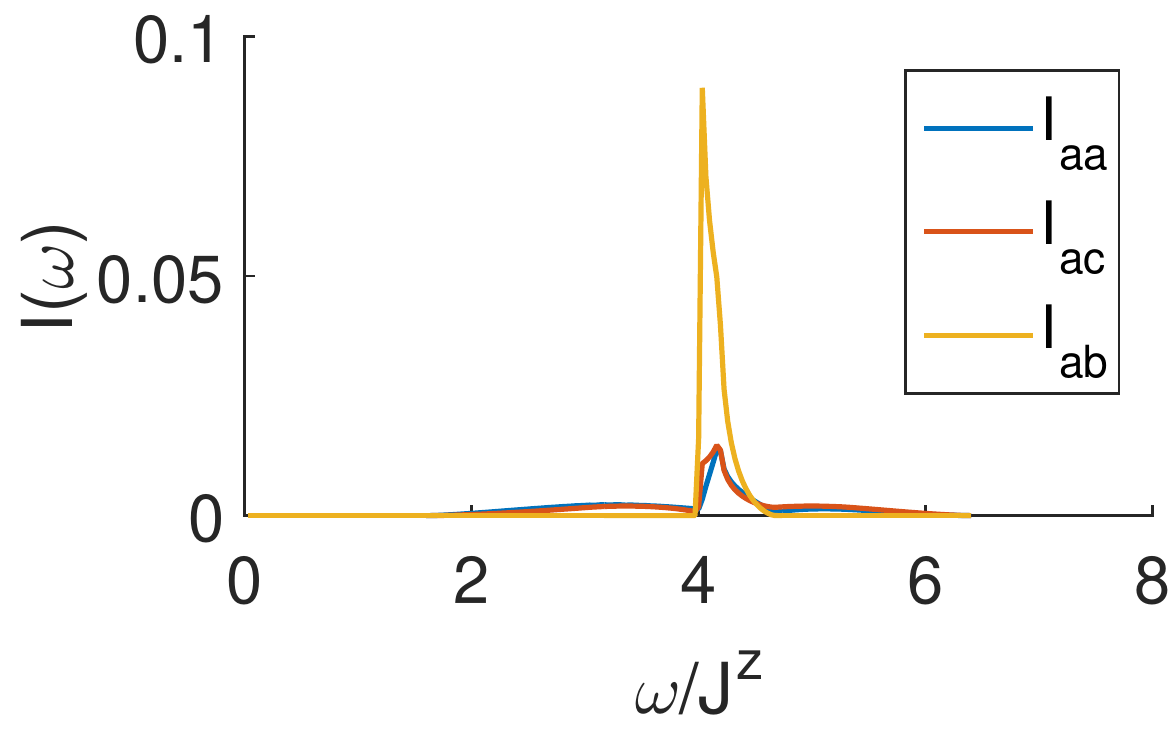}};
 			\node[anchor=south west,inner sep=0] (image) at (1.4,3.17) {(c)};
	 			\end{tikzpicture}
	 		\end{minipage}
 			\caption{(Color online) Raman intensities for the \Hs{0} lattice computed for: \mbox{(a) $J^x = J^y = J^z$ (gapless)}, (b) \mbox{$J^x = J^y = 1.43J^z$} (gapless), (c) \mbox{$J^x = J^y = 0.3J^z$} (gapped). The spectrum looks qualitatively the same throughout the gapless phase, in contrast with the 2D case in which the symmetric coupling point has extra symmetry. The maxima and minima in $I(\omega)$ originate from van Hove singularities and band edges, respectively,  for each band (see Fig. \ref{Spectrum}). The only two particle combination that couples to the $B_{1g}$ channel is one from the lower band and one from the upper band causing the spectrum $I_{ab}$ to vanish unless $4J^z < \omega < 4 \sqrt{5}J^z$.\cite{bandfoot}}
 			\label{H0}
 		\end{figure*}
 		
 	When we take $J^z \ne J^x = J^y$, the rotation symmetry is broken and there are four independent non-zero spectra.  Nevertheless, they are reduced to two because the LF relationship forces the three terms in the $A_{g}$-channel to be degenerate ($I_{xx} = I_{yy} = - I_{xx,yy}$). The two remaining spectra are represented by $I_{xx}$ and $I_{xy}$ of the $A_{1g}$ and $B_{1g}$ irreps. These are plotted for a gapless point $(J^x = J^y = 1.43 J^z)$ and a gapped point $(J^x = J^y = 0.3 J^z)$ in Figures \ref{2D}(b) and \ref{2D}(c). The breaking of the LF relationship is measured by a nonzero sum $I_{xx} + I_{xx,yy}$ in all cases. 
	 	
 	Along with the excitation gap seen in the gapped phase, the computed intensities show broad humps that are qualitatively similar to the isotropic case. The relative total spectral weight between the two active channels is plotted in Fig. \ref{weights} where we find that the relative weight is determined by the coupling anisotropy $J^x/J^z$. One can understand this dependence by considering the following limits: (1) $J^x/J^z = 0$, where the states are zero-dimensional, with no Raman response; and (2) $J^x / J^z \to \infty$, where the states are one-dimensional, with Raman intensity only in the $A_{2g}$ channel.
	 	
 	Apart from the broad hump there are also fine features related to the \ms density of states (shown in the left insets). In particular, the van Hove singularities of the saddle points in the dispersion lead to singularities in the derivative of the Raman response (shown in the right insets of Fig. \ref{2D}).

 	\subsection{3D \HH spectra}
 	
 		\begin{figure*}[htb]
 			\centering
 \includegraphics{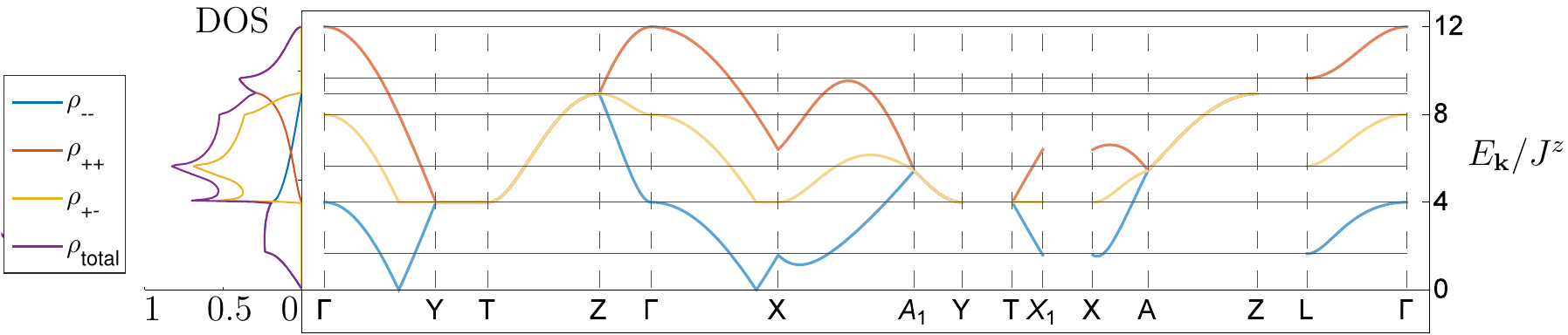}
 			\caption{(Color online) The \ms spectrum on the \Hs{0} lattice plotted along high symmetry lines. The inset on the left   shows the 2-DOS defined in Eq.~(\ref{2PartDOS}). For details of the  BZ see Appendix \ref{zones}. The two complex fermions per unit cell make two bands, $\varepsilon_{\pm,\mathbf{k}}$.  This gives three distinct configurations for the two-particle states of interest in Raman scattering: two with both excitations in the same band, and one with one excitation per  band. The 2-DOS is split into the contributions from each of the configurations
by $\rho_{mn}(\omega) = \sum_{\mathbf{k}} \delta(\omega - \varepsilon_{m,{\mathbf{k}}} - \varepsilon_{n,{\mathbf{k}}})$, where indices $m,n=\pm$. The maxima and minima in $I(\omega)$ appear as extrema in the spectrum of the appropriate band at some symmetry-distinguished point in the BZ.}
 			\label{Spectrum}
 		\end{figure*}
 	
 	Many of the characteristic features of the 2D Raman spectra carry over to the 3D lattices.
 Notably, because Raman scattering couples only to the \mssns, the spectra are broad, reflecting the total bandwidth\cite{widthfoot} $4z J^{av} = 4 (J^x + J^y + J^z)$ accessible to two-particle \ms excitations above the ground state, where $z$ is the number of neighboring sites.  
We note that this is also the energy range in which we would find magnons (spin-waves) if the state were ordered. Moreover, in the gapless spin liquid phases
we always find a linear Raman spectrum at low energies, as can be anticipated from the 1D nature of the fermi-surface (a line) and the linear dispersion away from that line. \cite{Schaffer,Hermanns}   
This feature, together with a gapped dynamical spin structure factor, are common to all of the \hh lattices. 

However, there are also a number of notable differences. 
The most striking one is the greater number of independent, non-vanishing spectra as a function of polarization, due to the lower symmetry and larger number of ways for light to couple to the lattice. 
This can be clearly seen in the Raman spectrum of the \Hs{0} lattice, shown in Fig. \ref{H0}. 
In particular, we find that $2I_{ac} = I_{bc}$ so that the $B_{2g}$ and $B_{3g}$ channels have identical response. Although no space group symmetries relate the $\mathbf{a}$ and $\mathbf{b}$ directions, this result can be understood in terms of the symmetry of an equivalent lattice model. The connectivity is unchanged by scaling $a \to \sqrt{2}a$, which leads to a lattice with screw symmetry composed of $C_4$ rotation about an inversion center, along with $a_3/2$ translation. Equivalently, the original lattices are symmetric under the scaling $a \to \sqrt{2}a, b \to b/\sqrt{2}$ followed by the same screw rotation.  This symmetry exchanges $\mathbf{a}$ and $\mathbf{b}$ as desired, without affecting the physics of the spin model. (We use this transformation only as a theoretical tool: in practice, macroscopically compressing the Li$_2$IrO$_3$ materials would break the symmetry between the spin-orbit distinguished directions -- see Refs. \onlinecite{Takayama} and \onlinecite{Modic} for material details). The factor of $2$ between the intensities $I_{ac}$ and $I_{bc}$ comes from the different projections along the actual $a$ and $b$ directions. 
In addition, the equalities $4I_{aa} = I_{bb}$ and $2I_{aa,cc} = I_{bb,cc}$ mentioned in the previous section are also guaranteed by this symmetry, leaving only two relationships that depend on the Loudon-Fleury form of the operator: $I_{cc} = 9I_{aa} = -3I_{aa,cc}$.  

\begin{figure}[htb]
 		\centering
 			\includegraphics[width=.9\linewidth, trim = 0mm 0mm 0mm 0mm, clip]{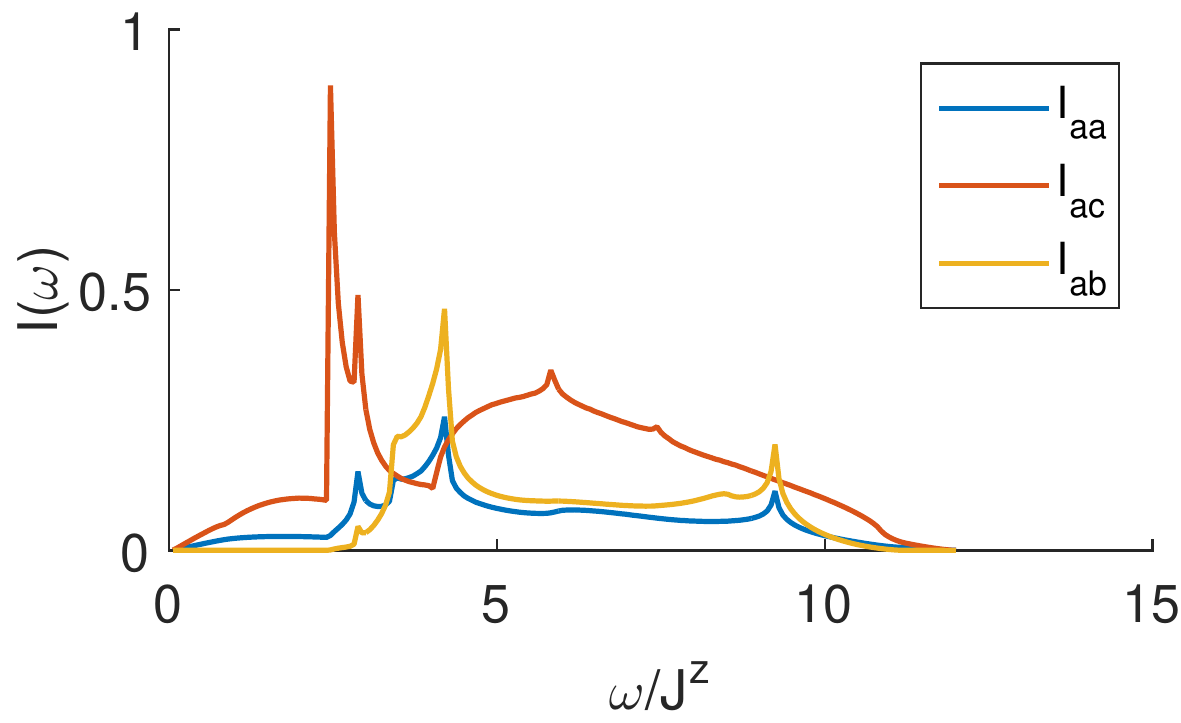} 
 		\caption{(Color online) Raman intensities for the \Hs{1} lattice in the zero-flux state at the isotropic point ($J^x = J^y = J^z$).}
 		\label{H1}
 	\end{figure}

\begin{figure}[htb]
 		\centering
 			\includegraphics[width=.9\linewidth, trim = 0mm 0mm 0mm 0mm, clip]{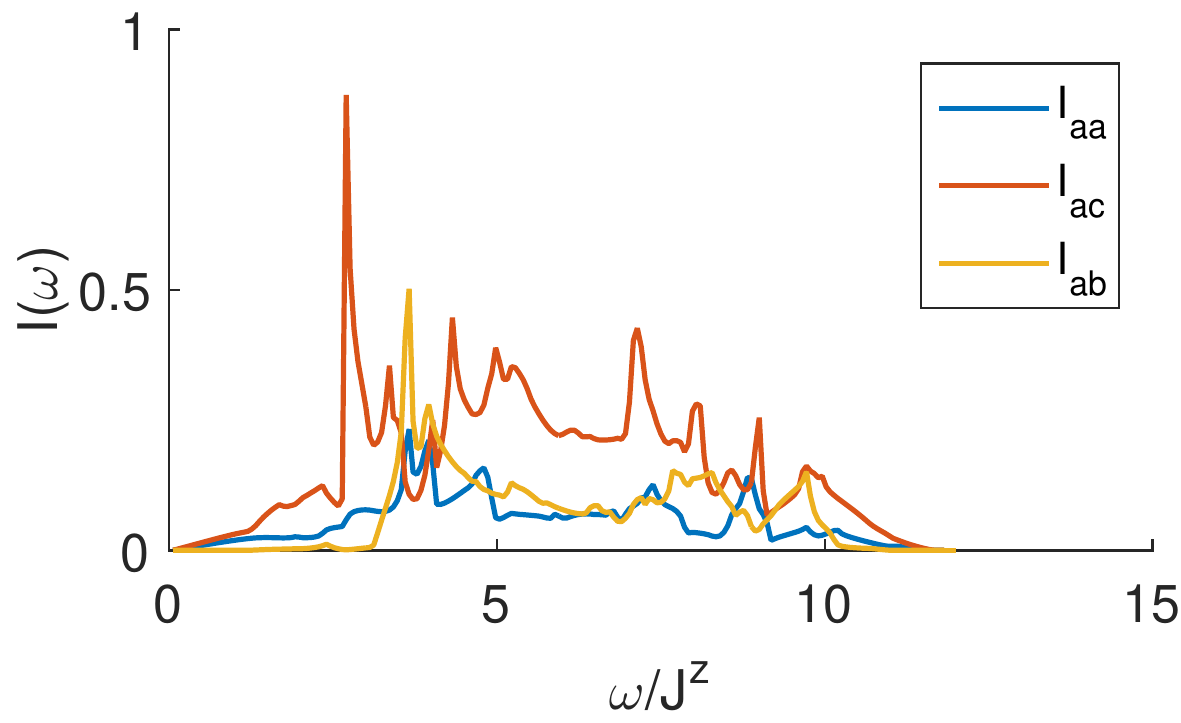} 
 		\caption{(Color online) Raman intensities for the H-1 lattice $\pi$-flux state at the isotropic point ($J^x = J^y = J^z$).\cite{Schaffer} }
 		\label{H1pi}
 	\end{figure}

Unlike in the 2D case, the low-frequency response for the \Hs{0} lattice shows a strong polarization dependence, with only one of the three polarization combinations showing a significant low-frequency response in the gapless phase.  
In fact, a strong polarization dependence in the low-energy response is common to all of the 3D lattices \Hs{n}, $n<\infty$:  the $B_{1g}$ channel $(I_{ab}$ in the Figure) is always inactive at low energy. This can be understood by considering the symmetry of the Raman operator in this channel. 
 There is a $\zt$ glide plane symmetry (see Appendix \ref{lattice symmetries}) common to all finite $n$ lattices.  
 Therefore the bands can be labeled even or odd under this symmetry. For momentum-dependent symmetries such as a glide plane, the parity of the Raman operators can be carried by momentum-dependent coefficients rather than the excitations. However, the Raman operator in the $B_{1g}$ channel ($R_{ab}$) is odd under the 
 glide plane transformation, and, since this operator does not vanish at $\mathbf{k}=0$, this parity must be carried by the \ms excitations that it creates. Therefore the pair of excitations must change the glide plane symmetry quantum number of the state it acts on.  Because glide plane symmetry is a $\zt$ symmetry, creating two \ms excitations in the same band is always a glide-even operation. Therefore the Raman operator in the $B_{1g}$  channel can only couple to two-particle states involving excitations from different bands.  In particular it cannot excite a pair of excitations both in the lowest band.  (Nor can it excite a pair in the highest band). This is apparent in Fig. \ref{H0}(a) where  $I_{ab}$ (the Raman intensity in the $B_{1g}$ channel) is non-vanishing only in the region $4J^z < \omega < 4 \sqrt{5}J^z$ -- i.e. only at energies accessible by exciting one \ms from each band.\cite{bandfoot} 
 For identification of the contributions of each band see Fig. \ref{Spectrum} below.  
  Since for general $n$ the \ms ground state contains only one gapless band, the spectrum $I_{ab}$ is gapped for all of the 3D lattices. See Figures \ref{H1} and \ref{H1pi} for similar vanishing of $I_{ab}$ on the \Hs{1} lattice. This is a particular result of the interplay between the fermionic fractionalization in the Kitaev spin liquid and the symmetries of the 3D \hh lattices.

 	Another notable difference between the Raman spectra of the \Hs{0} and \Hs{\infty} lattices is the number of maxima and minima.   The 2D lattice shows only a broad continuum (at least in the pure Kitaev model\cite{Perkins}), with only one broad peak.  In contrast, the 3D lattices show a number of rather sharp peaks in addition to the overall broad response. 
	This is even more noticeable in the spectrum of the \Hs{1} lattice, shown for the zero flux ground state in Fig. \ref{H1}, and for 
the $\pi$ flux ground state, which has an enlarged unit cell, in Fig. \ref{H1pi}, for $J^x = J^y = J^z$.  
	
	These new peaks can be understood by considering the momentum-locked two-particle density of states (2-DOS) defined in Eq.~(\ref{2PartDOS}), which the Raman spectrum closely emulates. For the $\mac{H}$-$0$ lattice, this is plotted in Fig. \ref{Spectrum}, along with the band structure along high symmetry lines.  
The 4-site \ms unit cell implies that there are two \ms bands, and therefore three types of quasiparticle pairs above the ground state. As the figure makes clear, the sharp features correspond either to band edges, or to van Hove (VH) singularities that occur near the high-symmetry points in the Brillouin-zone. 

	 \begin{figure*}[htb]
	 	\centering
	 	\begin{minipage}{.32\linewidth} 
		 	\begin{tikzpicture}
	 	\node[anchor=south west,inner sep=0] (image) at (0,0) {
	 	\includegraphics[width=\linewidth, trim = 0mm 0mm 0mm 0mm, clip]{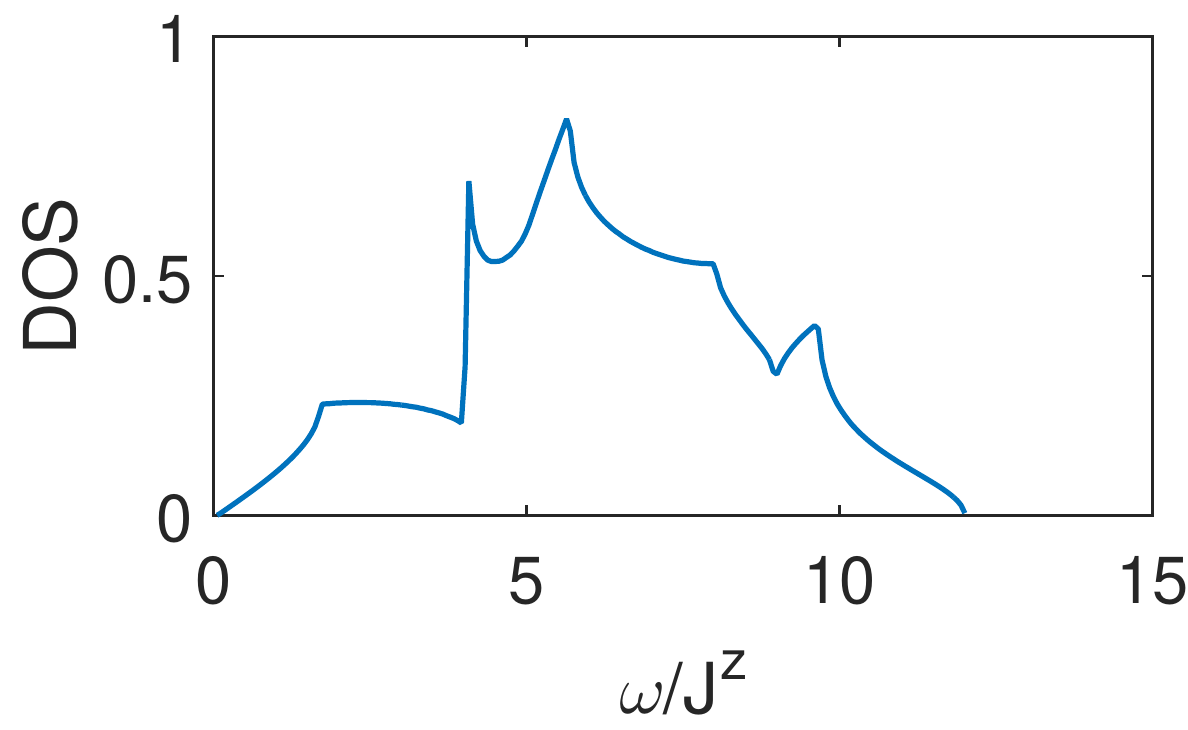} }; 
		 \node[anchor=south west,inner sep=0] (image) at (1.2,2.8) {(a)};
		\end{tikzpicture}
	\end{minipage} 
	\begin{minipage}{.32\linewidth}
		\begin{tikzpicture}
	\node[anchor=south west,inner sep=0] (image) at (0,0) {
	 	\includegraphics[width=\linewidth, trim = 0mm 0mm 0mm 0mm, clip]{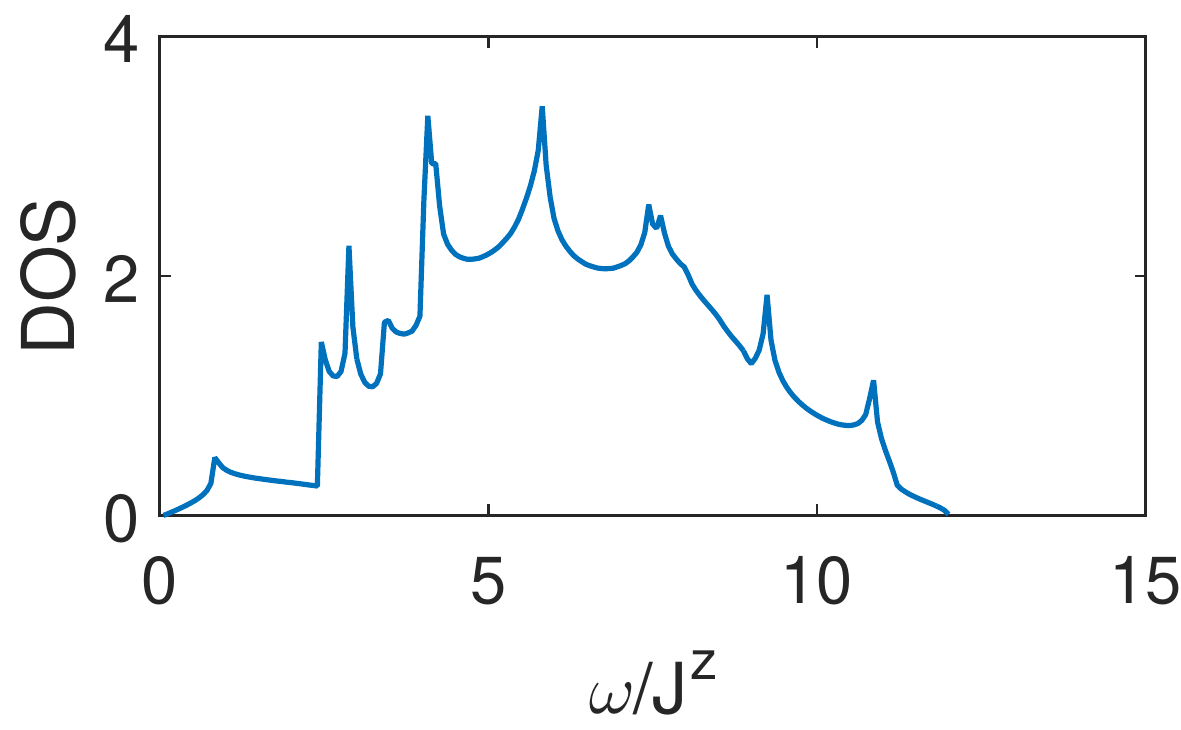}}; 
	 \node[anchor=south west,inner sep=0] (image) at (0.95,2.8) {(b)};
	 \end{tikzpicture}
	\end{minipage} 
	\begin{minipage}{.32\linewidth}
		\begin{tikzpicture}
		\node[anchor=south west,inner sep=0] (image) at (0,0) {
	 	\includegraphics[width=\linewidth, trim = 0mm 0mm 0mm 0mm, clip]{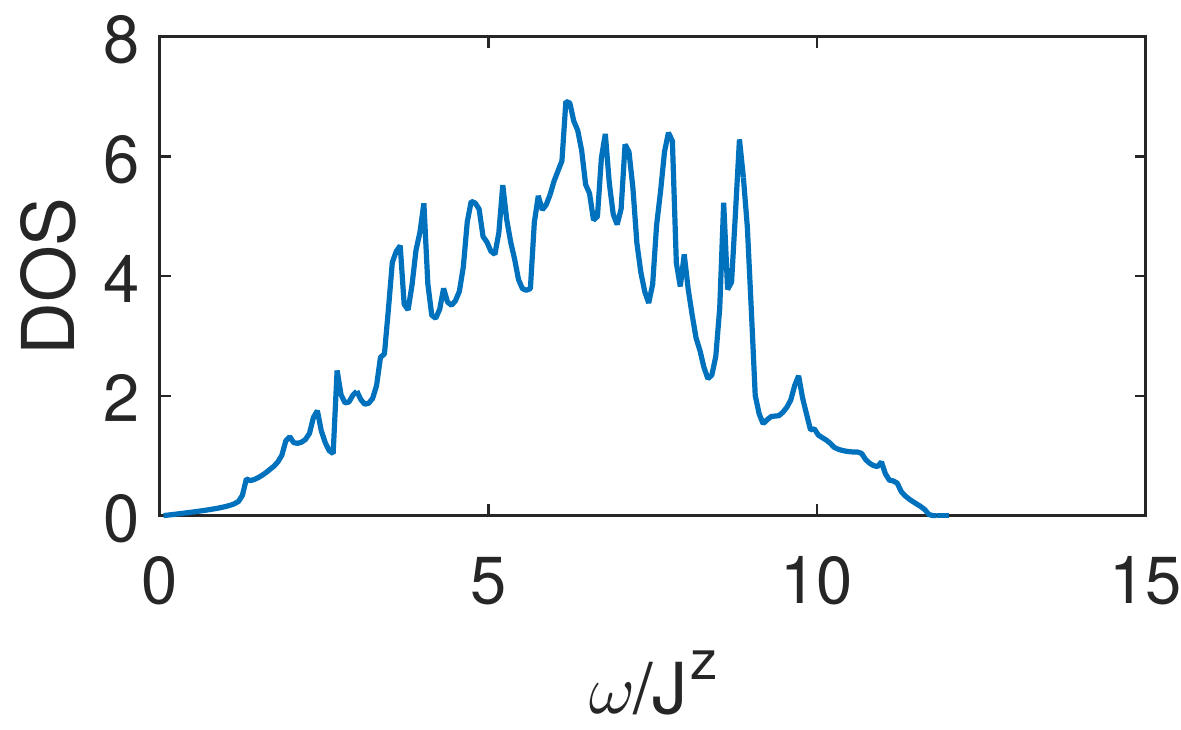}}; 
	 \node[anchor=south west,inner sep=0] (image) at (0.95,2.8) {(c)};
	 \end{tikzpicture}
	 	 \end{minipage}
	 	\caption{2-DOS for the \mss on the (a)  \Hs{0}, (b)  \Hs{1} and (c) \Hs{1}$\pi$  lattices. Up to polarization dependence these spectra determine the character of the Raman intensities. The number of maxima and minima grows linearly with the number of combinations of bands for two-particle states, which increases from left to right due to increasing unit cell size.}
	 	\label{DOSs}
	 \end{figure*}
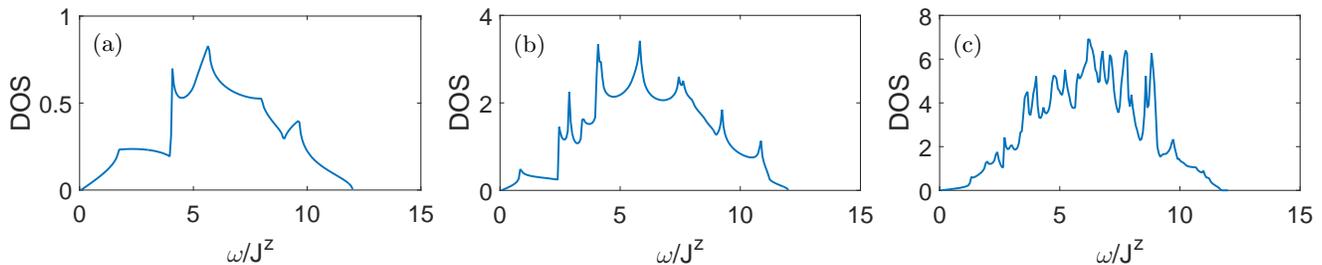

Fig. \ref{DOSs}  compares the 2-DOS of the  $\mac{H}$-$0$,  $\mac{H}$-$1$, and  $\mac{H}$-$1 \pi$ lattices.  (Here  $\mac{H}$-$1 \pi$ refers to the $\pi$ flux ground state of the  $\mac{H}$-$1 $ lattice, while  $\mac{H}$-$1$ denotes the energetically proximate 0-flux state.)
 Since the number of sites in the unit cell increases with $n$, so does the number of sharp features in the Raman spectrum (see Figs. \ref{2D}a, \ref{H0}, \ref{H1}, and \ref{H1pi}).  In the 2D lattice there is only one \ms band, and the Raman spectrum has only an upper edge, a lower edge and a VH singularity.  For the \Hs{n} lattice there are  $n+2$ bands. 
 Then each pair of \mss has a maximum and a minimum, as well as possibly one or more VH singularities, leading to a number of features growing approximately quadratically in the unit cell size. 

 		\begin{figure}[htb]
 			\centering
 			\begin{tikzpicture}
 			\node[anchor=south west,inner sep=0] (image) at (0,0) {
 				\includegraphics[width=.9\linewidth, trim = 0mm 0mm 0mm 0mm, clip]{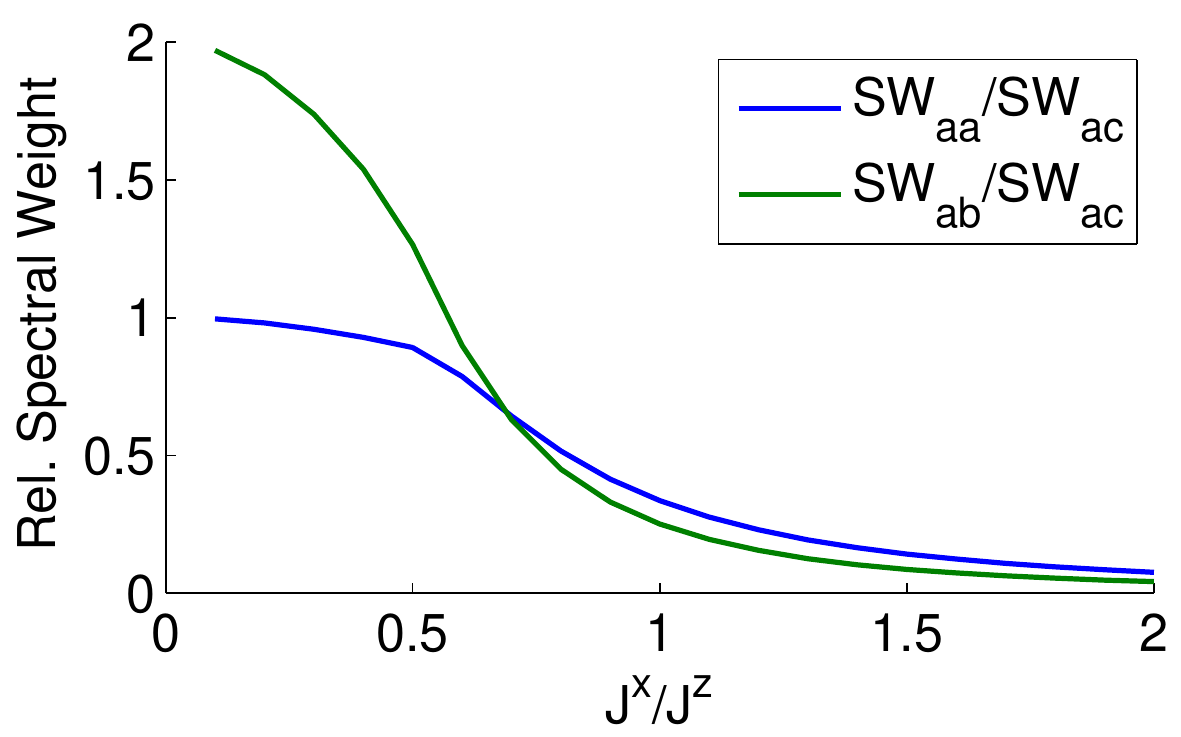} };
 			\end{tikzpicture}
 			\caption{(Color online) Relative spectral weights for Raman on the \Hs{0} lattice as a function of the bond-coupling anisotropy. }
 			\label{weights2}
 		\end{figure}

 To summarize, the Raman spectra of the 3D \hh Kitaev models differ from those of the 2D model in both their polarization dependence and the number of sharp peaks.  However, a number of key features are common to all of these spin liquid phases, including the bandwidth and the fact that symmetry and LF relations determine the polarization dependence. Moreover, in the 3D lattices the polarization dependence is again characteristic of the coupling anisotropy $J^x/J^z$.  This is illustrated in Fig. \ref{weights2}, which shows the relative 
 total spectral weights (integrated over all frequency) of the representative polarization combinations for the \Hs{0} lattice. We have not plotted the analogous weights for the \Hs{1} lattice because the results are nearly indistinguishable. The limits of $J^x/J^z = 0 , \infty$ can be understood the same way as for the 2 lattice. For the 3D lattices we do not have polarization independent response at the symmetric coupling point $J^x = J^y = J^z$ as there is no change in symmetry at that point.

\subsection{Discussion}

Our discussion so far has focused on the exactly solvable Kitaev Hamiltonian at zero temperature.  
While this has the advantage of allowing us to evaluate the Raman spectrum exactly, in any real experiment additional spin-spin couplings would be present, which would have to be taken into account to predict the Raman response even within the spin liquid phase.  
In addition, it is useful to understand how finite temperature affects the spectra shown above.  Here we will comment on how our results are expected to change at finite temperatures, and in the presence of perturbations away from the pure Kitaev Hamiltonian. 
 Our discussion is qualitative rather than quantitative in nature, due to the technical challenges involved in repeating our calculation in these regimes.  

First, we consider how the presence of additional spin-spin couplings would affect our result.  In general, such interactions will not commute with the conserved quantity $W_P$ (Eq.~(\ref{W})). As a result there will be some flux fluctuations in the ground state, which is no longer exactly described by a \ms band structure. However, the most noticeable change is likely the one due to the change in the LF operator, which will have some finite probability to excite fluxes. 
 Ref. \onlinecite{Perkins}  showed, using a perturbative treatment of a small nearest-neighbor Heisenberg interaction, that this leads to an additional peak in the Raman response at the energy to create four flux excitations.  For the 3D lattices we expect the situation to be qualitatively similar: adding a small Heisenberg (or other symmetry-preserving) perturbation to the Kitaev Hamiltonian should lead to a new peak at the energy scale of the gap for two flux-loop excitations.  
 We note that quantitatively this flux gap for the 3D lattices is expected to be smaller with respect to the Kitaev exchange coupling than in the 2D case. \cite{infinite-D}
  
As discussed in Section \ref{PolSec}, the polarization dependence is derived from two considerations.  Many of the relationships between polarizations are fixed by lattice symmetry alone.  However, 
some of the relationships between different polarization channels found in Appendix \ref{pol} come from the close relationship between the Hamiltonian and the LF Raman operator, which we call the LF relationship. 
In doing perturbation theory, it is therefore essential to treat the Hamiltonian and the Raman operator to the same order to preserve this relationship. We note that Ref. \onlinecite{Perkins} treated corrections to the Raman operator but not to the Hamiltonian (and therefore the ground state) in the presence of a Heisenberg perturbation.  They consequently found that this term breaks the polarization independence described above. 
However, adding perturbative corrections to the same order in the ground state should restore this symmetry, and produce the polarization-independent Raman response anticipated from the combination of symmetry and the LF relationship. 

The LF relationships reported here are expected to hold for all nearest-neighbor spin-exchange Hamiltonians. Breaking an LF relationship results in more linearly-independent spectra. On the 2D lattice the same LF relations apply for any bond orientation. However, for the 3D lattices the LF relationships can be broken by the introduction of further neighbor spin-exchange terms in the Hamiltonian. See Appendix \ref{pol} for more details. There are also ways to break these relations that apply to all the lattices. First, perturbations that do not come from an electron virtual hopping process -- such as a time-reversal (TR) breaking magnetic field -- would break this relationship. 
Also, tuning the incoming photon energy into the resonant regime (near the Mott gap) will allow multiple electron hoppings and therefore leads to the appearance of higher order spin terms in the Raman operator that are not present in the effective Hamiltonian.\cite{Ko,Shastry} This is also expected to break the LF relationship.  
We also note that a small TR-breaking perturbation can also lead to a change in the power law of the low energy \ms DOS.\cite{Hermanns} We will discuss this in the context of Raman scattering in future work.\cite{us}  

Next, let us consider the effects of finite temperature.  At temperatures much lower than the flux gap, flux excitations are rare, and the spectrum  is expected to look very similar to the zero-temperature result derived here. The transition out of the spin-liquid phase occurs when flux loops proliferate, confining the dispersing \mssns. The transition is expected to be different in the 2D and 3D cases, due to the different nature of the flux excitations. In 2D the flux defects are point-like, with no long-ranged flux-binding interactions. At any finite temperature there will therefore be a finite density of {\it unbound} fluxes, which immediately confines the \mss at long length scales due to the mutual semionic statistics between \mss and fluxes.  In 3D, however, the flux defects are loops, with an energy cost proportional to the total loop length $E \propto \alpha L$, where length $L$ is measured in plaquettes. Below $T_c$, which is set by $\alpha$,\cite{infinite-D} the flux loops stay small, and since in 3D the mutual statistics is only felt by \mss when they go \emph{through} a loop, a finite density of small loops does not lead to confinement. 
At temperatures above $T_c$ flux loops may still be rare but the entropy gain in having larger loops wins over the energy cost and there is a small density of loops of arbitrary length.\cite{infinite-D,Nasu}

Above the critical temperature $T_c$ there is no spin liquid phase because the \mss are confined degrees of freedom and therefore cannot form bands. However, there is a confinement length scale (set by the density of loops) below which the \mss can still propagate. This sets an energy scale above which the \mss may have dynamics that reflect the deconfined (spin liquid) state. The finite temperature transition has been considered in quantum Monte-Carlo for the \Hs{0} lattice.\cite{Nasu2,Nasu,Nasu3} The transition temperature turns out to be about $T_c=0.011 T_K$ for the \Hs{0} lattice with $J^x = J^y = J^z$, where $T_K$ is the temperature corresponding to the Kitaev exchange coupling $J^\alpha$. In fact, Ref. \onlinecite{Nasu3} claims that the dominant changes appear only in the low-energy \ms DOS  as the temperature increases in the range $T_c < T < T_{K}$.  Therefore,  the high energy features of our Raman response could be observable up to $\lesssim T_K$, which is also consistent with  recent experiments. \cite{Sandilands}

Finally, let us speculate on the possibility of detecting spin-liquid signatures in the Raman spectra of the A$_2$IrO$_3$ \hh irridates, and $\alpha$-RuCl$_3$.  Experimentally, these materials have been found to order at low temperatures.\cite{Kimchi3} This is not unexpected, since the appropriate spin Hamiltonian contains both Kitaev interactions and other spin-exchange terms.\cite{Kimchi,Sizyuk,Rousochatzakis2015}  However, one might hope that signatures of a proximate spin-liquid phase are observable at temperatures above the ordering temperature $T_N$.  Specifically, if the quantum transition out of the spin liquid phase due to perturbations is first order, finite temperatures fluctuations for $T> T_N$ will carry signatures of the spin liquid state.  
This case has already been made in Ref. \onlinecite{Alpichshev} -- see also Ref. \onlinecite{Balents}. 


It is therefore not unreasonable to expect that the Raman spectra of the currently available compounds in the temperature range $T_N < T < T_K$ 
could display high-energy features of the Kitaev spin-liquid state. For instance, after isolating the magnetic contributions from the ones due to phonons, one could look for the band-resolved polarization dependence found for the 3D lattices above. As already mentioned, much of the polarization dependence is dictated by symmetry for any spin-exchange Hamiltonian whose eigenstates respect the lattice symmetry. Overall, we believe that any quantitative agreement with the high-energy
Raman spectra reported here as well as with the distinct polarization
dependence indicative of QSL phases on the harmonic
honeycomb lattices would make the case for these long sought-after states
of matter much stronger.

In summary, we have studied Raman scattering in the Kitaev spin model on the \hh lattices.  We find that the Raman response of the spin liquid has the following distinctive features:
\bit
\m A broad continuum response determined by the \ms 2-DOS and specific features characteristic of the unit cell size.
\m A gapless phase with a low energy asymptotic power law reflecting the exotic Fermi surfaces.
\m Linear relations between distinct polarization channels dictated by the LF relations.
\m Band-resolved polarization dependence in certain channels for the 3D lattices that reflects the particular fractionalization of the Kitaev spin liquid.
\m Polarization dependence of the spectral weights that is related to the exchange-coupling anisotropy. 
\eit

	\section*{Acknowledgements}
	We acknowledge useful discussions with Yuriy Sizyuk, Itamar Kimchi, Alex Edelman, James Analytis, Ioannis Rousochatzakis, Gia-Wei Chern, Roderich Moessner, Dima Kovrizhin, Phillip Gegenwart, Cenke Xu, Max Metlitski, Leon Balents, and Peter Lemmens. FJB is supported by NSF DMR-1352271 and Sloan FG-2015-65927.  The work of JK is supported by a Fellowship within the Postdoc-Program of the German Academic Exchange Service (DAAD). NP acknowledges the support from NSF DMR-1511768.

	\appendix
	
	\section{\HH lattices}\label{sec:lattice}
		 
Here we provide additional details on the structures, symmetries, and Brillouin zones of the \hh lattices.  

We first introduce some vocabulary.  On each lattice the $x$ and $y$ bonds 
make chains along two distinct directions labeled $A$ and $B$ in Eq.~(\ref{3Dds}). We call the $z$ bonds that connect parallel chains \emph{rungs} and the ones connecting opposite chains \emph{bridges} -- see Fig. \ref{fig:Harmonic_Lattices}. The 2D honeycomb lattice has only rungs making up a single ladder. In the $c$-direction, the \Hs{n} ladders have $n+1$ chains connected by $n$ sets of rungs before a bridge to the opposite ladder.

For the 3D lattices, the coordinates that we use are given in the main text.  (See Eqs.~(\ref{LatticeVectorsEq}) and~(\ref{3Dds})).  
	For the \Hs{\infty} lattice (2D honeycomb) it is more convenient to use coordinates in the lattice plane as in Ref. \onlinecite{Perkins}. For this case we rename $c$ as $y$ and define $x = \hat{\mathbf{B}} \propto (a,b,0)$. We reserve square brackets for vectors in these coordinates. The unit vectors are $n_{1/2} = [3,\pm \sqrt{3}]/2$ and the bond vectors (from odd sublattice to even) are given by
	\begin{align}\label{2Dds}
	d^z &= [0,1]  &(\textrm{blue}) \nonumber \\
	d^{x} &= \frac{1}{2}[-\sqrt{3}, -1]  &(\textrm{red}) \nonumber \\
	d^{y} &= \frac{1}{2}[\sqrt{3}, -1]  &(\textrm{green}).
	\end{align}

	\subsection{Lattice symmetries}\label{lattice symmetries}
	
	Here we review the symmetries of the three lattices, which are used in Appendix \ref{pol} to constrain the polarization dependence, and also below to constrain the ground state flux configuration. 	
	
	The $n<\infty$ \Hs{n} lattices all have C$_2$ symmetries in the orthogonal $\hat{\mathbf{a}}, \hat{\mathbf{b}},$ and $\hat{\mathbf{c}}$ directions about the center-points of bridges. Note that the $a$ and $c$ ones swap $x$ and $y$ bonds, so that if $J^x \ne J^y$ the point group would be broken down to $C_{2h}$, although we do not consider that case here. They also have inversion centers at the midpoint between the center-points of two near bridges. All of them have glide planes (reflection $\times$ translation by half a unit vector) in all three of the directions $\mathbf{a},\mathbf{b},\mathbf{c}$. For the odd-$n$ lattices, the glide planes are $a,b,$ or $c$-reflection about a bridge-center with $a_3/2$ translation. The odd-$n$ lattices also have mirror symmetries in the $c$-direction about the inversion centers.

	For the even-$n$ lattices the glide planes' reflections pass through the inversion centers, which sit on the center of an $x$ or $y$ bond at the midpoint between two bridge centers. The glide plane $c\to-c$ must be composed with $a_2$-translation or $a_1$-translation for centers of $A$- or $B$-type bonds respectively. The glide plane about $a$ on $J^x_A$ and $J^y_B$ bonds is composed with $(a_3 + a_1)/2$ translation while on $J^y_A$ and $J^x_B$ bonds it includes $(a_3 + a_2)/2$ translation. Finally, the glide plane taking $b \to -b$ includes $a_3/2$-translation or $(a_1+a_2+a_3)/2$ translation when the reflection plane passes through $x$- or $y$-type bonds respectively. 

	The 2D honeycomb lattice is more constrained by symmetries, including $C_6$ rotation and mirror planes.\cite{You} The 3D point group of the 2D lattice is $D_{3d}$. The projection of the point group onto the 2D plane is generated by $\{C_6,\sigma\}$ where 
	\begin{align}\label{rep2}
	C_6 =  
	\left(\begin{array}{cc}
	1 & -\sqrt{3} \\ \sqrt{3} & 1
	\end{array}\right)
	\hspace{1cm}
	\sigma = \left(\begin{array}{cc}
	-1 & 0 \\ 0 & 1
	\end{array}\right).
	\end{align} 	 
	
	\subsection{Constraining the ground state flux}
Because the mirror symmetries on the odd-$n$ lattices do not pass through any lattice points we can use Lieb's theorem\cite{Lieb} to constrain the ground state flux through the plaquettes through which the mirror plane passes.
 On the \Hs{1} lattice this forces the hexagons as well as the symmetric $14$-site plaquettes to carry $0$-flux. 
 The ``$\pi$-flux state" fulfills these requirements.\cite{Schaffer} That flux configuration can be realized by changing the sign of $u_{\left<ij\right>^\alpha}$ on every other bridge bond or by threading $\pi$ flux through every rhombus in a projection along the $c$-direction. 
 However, the gauge illustrated in Fig. \ref{fig:Harmonic_Lattices} (with a sign change on every other $x$ and $y$ bond in the $a_1$ and $a_2$ directions on one ladder type) is most convenient because it accommodates the $\pi$ flux state with the smallest possible unit cell (i.e. double the unit cell of the $\mac{H}$-$1$ lattice).

	\subsection{Brillouin zones}\label{zones}
	
	The even-$n$ lattices are face-centered cubics (Fddd), while the odd-$n$ are base-centered (Cccm).\cite{Lee3,Schaffer,Takayama,Modic} The 2D honeycomb lattice has point group $D_{3d}$, which is distinct from the others. The three lattices considered here are representatives of these three space groups. 	

	For the \Hs{0} lattice our choice of BZ is ORCF$_1$ from table 9 of Ref. \onlinecite{cell}. 
	It can be parametrized by \cite{cell}
	\begin{align}
	|k_c|<\frac{\pi}{3}, \hspace{.5 cm} |k_b|<\frac{\pi}{\sqrt{2}}, \hspace{.5 cm} |k_a| < \frac{29 \pi}{36} - \frac{|k_b|}{\sqrt{2}} - \frac{|k_c|}{3}. 
	\end{align} 
	In coordinates $(k_a,k_b,k_c)$, the high-symmetry points used in Fig. \ref{Spectrum} are
		\begin{align}
		\Gamma &= (0,0,0) \\
		\begin{split} \nonumber
		A &= \left( \frac{25 \pi }{36}, 0 , \frac{\pi }{3} \right) \nonumber \\
		A_1 &= \left(\frac{11 \pi }{36},\frac{\pi }{\sqrt{2}}, 0\right) \nonumber \\
		L &= \left(\frac{\pi }{2},\frac{\pi }{2\sqrt{2}},\frac{\pi }{6}\right) \nonumber \\
		T &= \left(0,\frac{\pi }{\sqrt{2}},\frac{\pi }{3}\right) \nonumber
		\end{split}\quad \quad
		\begin{split}
		X &= \left(\frac{29 \pi }{36},0,0\right) \nonumber \\
		X_1 &= \left(\frac{7 \pi }{36},\frac{\pi }{\sqrt{2}},\frac{\pi }{3}\right) \nonumber \\
		Y &= \left(0,\frac{\pi }{\sqrt{2}},\frac{\pi }{3}\right) \\
		Z &= \left(0,0,\frac{\pi }{3}\right).
		\end{split}
		\end{align}
			
	For the \Hs{1} lattice we use the ORCC Brillouin zone from Ref. \onlinecite{cell}.
	\begin{align}
	|k_c| < \frac{\pi}{6}, \hspace{.5cm} |k_b| < \frac{\pi}{\sqrt{2}}, \hspace{.5cm} |k_a|< \frac{3\pi}{4} - \frac{|k_b|}{\sqrt{2}}.
	\end{align}
	These generalize quite simply to other even and odd-$n$ \hh lattices.
	
	For the 2D lattice we take a rectangular Brillouin zone.
	\begin{align}
	0 < \frac{2\pi}{3} - |k_{x}|/\sqrt{3} -|k_y|. 
	\end{align}

	\subsection{Hopping matrices}\label{Hopping}
	
	The \ms Hamiltonian on the \Hs{n} lattice in the flux background $\Phi$ is specified by the matrix $\mathcal{D}_{\mathbf{k}}^\eta$ from Eq.~(\ref{Hc}) (recall $\eta = n(\Phi)$). For each lattice the sites are numbered by their position in the positive $\mathbf{c}$-direction. We choose a gauge such that $u_{\left<ij\right>^\alpha} = +1$ if $i$ is at an odd site (yellow) and $j$ is at an even site (white) -- see Fig. \ref{fig:Harmonic_Lattices}.  The hopping matrices are \cite{Schaffer,Kitaev}
	\begin{align}
	\mathcal{D}_{\mathbf{k}}^{\infty(0)} = J^z + J^x e^{i\mathbf{n}_1 \cdot \mathbf{k}} + J^y e^{i\mathbf{n}_2 \mathbf{k}} \\
	\mathcal{D}_{\mathbf{k}}^{0(0)} = \left[\begin{array}{cc}
	J^z & A_\mathbf{k}^x e^{-ik_3} \\ B_\mathbf{k}^x & J^z
	\end{array}\right] \\
	\mathcal{D}_{\mathbf{k}}^{1(0)} = 
	\left[\begin{array}{cccc cccc}
	J^z & 0 & 0 & A_\mathbf{k}^y e^{-ik_3} \\
	A^x_\mathbf{k} & J^z & 0 & 0\\
	0&  B^y_\mathbf{k} & J^z &  0 \\
	0 & 0 & B^x_\mathbf{k} & J^z
	\end{array}\right]	 
	\end{align}
	\begin{align*}
	A^{x}_\mathbf{k} &= J^{x}_A + J^{y}_A e^{ik_1} \hspace{.4 cm} & B^{x}_\mathbf{k} &= J^{x}_B + J^{y}_B e^{ik_2} \\
	A^{y}_\mathbf{k} &= J^{y}_A + J^{x}_A e^{-ik_1} \hspace{.4 cm} & B^{y}_\mathbf{k} &= J^{y}_B + J^{x}_B e^{-ik_2}, \\
	\end{align*}
	where $k_i = \mathbf{k}\cdot\mathbf{a}_i$ and the subscript on the couplings distinguishes $J^{x}$ and $J^y$ bonds that lie on chains consisting of $A$ and $B$ bonds -- see Eq.~(\ref{3Dds}). The $\pi$-flux Hamiltonian requires a doubled unit cell. For a basis we take
	\begin{align}
	\mathbf{c}^{\mathsmaller T} = \left( c_1, c_3 , c_5, c_7, c_1', ... , c_7'; c_2,c_4,c_6,c_8,c_2',...,c_8' \right),
	\end{align}
	where we have dropped the $\mathbf{k}$-dependence to simplify the expression.
	In the gauge illustrated in Fig. \ref{fig:Harmonic_Lattices} (the sign of $u_{\left<ij\right>^\alpha}$ is switched on the circled bonds), with the unit cell doubled in the $a_2$ direction, the hopping is described by	
{\small			\begin{align*}
			\mathcal{D}_{\mathbf{k}}^{1(\pi)} = \hspace{7.5cm}  \nonumber\\
			\left[\begin{array}{cccc cccc}
			J^z & 0 & 0 & A^y_\mathbf{k} e^{-ik_3}      &     0 & 0 & 0 & 0\\
			A^x_\mathbf{k} & J^z & 0 & 0		&		0 & 0 & 0 & 0\\
			0&  J^y_B & J^z &  0 	&		0 & J^x_Be^{-2ik_2} & 0 & 0\\
			0 & 0 & J^x_B & J^z &      0 & 0 & J^y_B & 0 \\ 
			0 & 0 & 0 & 0	&	J^z & 0 & 0 & A^{y-}_\mathbf{k} e^{-ik_3} \\
			0 & 0 & 0 & 0				&	A^{x-}_\mathbf{k} & J^z & 0 & 0 	\\
			0 & J_B^x & 0 & 0				&	0&  J^y_B & J^z &  0	\\
			0 & 0 & J^y_B e^{2ik_2} & 0	&	0 & 0 & J^x_B  & J^z
			\end{array}\right] ,
			\end{align*}	}
		where we define
		\begin{align}
		A^{x-}_\mathbf{k} &= -J^{x}_A + J^{y}_B e^{ik_1} \hspace{.4 cm} & A^{y-}_\mathbf{k} &= -J^{y}_A + J^{x}_A e^{-ik_1}.
		\end{align}
	
	We have confirmed that the energy of the $\pi$-flux state is lower on the \Hs{1} lattice by studying the translation invariant system in the thermodynamic limit with an enlarged unit cell.\cite{Schaffer} In particular, at $J^x = J^y = J^z \equiv J$ the $0$-flux state has energy $-0.77397 J$ per site while the $\pi$-flux has $-0.77611 J$. For comparison, the local flux gap on the \Hs{0} lattice is $\sim 0.1J$. \cite{infinite-D}

\section{Explicit Raman correlation functions}\label{Intense}
Here we show the equivalence between Eqs.~(\ref{I-terms}) and~(\ref{result}) in the main text, provided that the Raman operator is of the form given in Eq.~(\ref{Rk}).   We consider the response of a quadratic fermionic Hamiltonian in the presence of a quadratic fermionic Raman operator. Assuming translation invariance, Eqs.~(\ref{I-terms}) and~(\ref{Rk}) can be written:
\begin{align*}
I(\omega) &= \int dt \braket{0| R \, e^{i(\omega-H+i\delta)t} R|0}, \\ 
R &= \sum_\mathbf{k} R_\mathbf{k}, \\
R_\mathbf{k} & =\left[ A_{mm',\mathbf{k}} a^\dagger_{m,\mathbf{k}}a_{m',\mathbf{k}} \right. \\  &\hspace{1cm} \left.+ \frac{1}{2}\left(B_{mm',\mathbf{k}} a^\dagger_{m,\mathbf{k}} a^\dagger_{m',-\mathbf{k}} +  h.c.\right)  \right]. \nonumber
\end{align*}
Inserting the Fourier decomposition for $R$ into the expression for the Raman intensity gives
\begin{align}\label{II}
I(\omega) &=  \int dt \sum_{\mathbf{k},\mathbf{k}'} e^{i\omega t} \braket{0|R_\mathbf{k}(t) R_{\mathbf{k}'}(0)|0}.
\end{align}
Notice that only the $B_{mm',\mathbf{k}}$ terms survive in (\ref{II}) at zero temperature since the $A_{mm',\mathbf{k}}$ terms annihilate the ground state. 
Therefore,
\begin{align*}
I(\omega) &= \sum_\mathbf{k} \frac{\pi}{2} \delta(\omega - \varepsilon_{m,\mathbf{k}} - \varepsilon_{n,\mathbf{k}}) \\ &\times \braket{0|\left(B^*_{mm',\mathbf{k}} a_{m',-\mathbf{k}} a_{m,\mathbf{k}}\right) \left(B_{nn',\mathbf{k}} a^\dagger_{n,\mathbf{k}} a^\dagger_{n',-\mathbf{k}}\right)|0}, 
\end{align*}
where we have used that $a^{\dagger}_{n,\mathbf{k}}(t)= e^{-it \varepsilon_{n,\mathbf{k}}}a^{\dagger}_{n,\mathbf{k}}(0)$ in the Heisenberg picture. Applying the anti-commutation relations and writing the sum over bands explicitly we find:
\begin{align}\label{result2}
I(\omega) &= \pi \sum_{\mathbf{k} \ne 0}\sum_{m,m'}  \delta(\omega - \varepsilon_{m,\mathbf{k}} - \varepsilon_{m',\mathbf{k}}) B^*_{mm',\mathbf{k}}B_{mm',\mathbf{k}},
\end{align}
The non-trivial commutation relations between $a^\dag_\mathbf{k}$ and $a^\dag_{-\mathbf{k}}$ for $\mathbf{k}=-\mathbf{k}$ lead a cancellation of the $\mathbf{k}=0$ term. Since the density of state vanishes at $\mathbf{k}=0$ the canceled term also vanishes in thermodynamic limit. This leads to Eq.~(\ref{result}) of the main text. 

Although the \ms Hamiltonian is translation-invariant in the flux sectors considered in the main text, the introduction of isolated fluxes would break this symmetry. To evaluate the Raman spectra at finite flux density requires numerical simulations on finite-sized lattices.  Though we have not carried these out, based on the numerics of Ref. \onlinecite{Nasu3} we expect that these translation-breaking terms will affect the qualitative conclusions given in the main text only at low frequencies.  

Finally, a few numerical details. The Brillouin zone integral implicit in (\ref{result}) is computed separately for each polarization combination. Numerically, we sample the Brilloiun zone and sort the points by energy. Then for a given $\omega$, the delta-function is replaced by a boxcar function of finite width. Therefore, the integral is taken as a sum over points whose energies satisfy the delta function constraint up to a small tolerance in energy. A similar integration method was used in Ref. \onlinecite{Ko}. Error estimates are obtained by randomly shifting the mesh and recomputing the results. 
Typical error estimates for the value of $I(\omega)$ at a given value of $\omega$ are $\lesssim 0.2 \%$.

	\section{Polarization dependence}
	\label{pol}
	
	As discussed in the main text, the Raman operator ${R} = \sum_{\alpha} (\boldsymbol{\epsilon}_{\textrm{in}} \cdot \mathbf{d}^\alpha) (\boldsymbol{\epsilon}_{\textrm{in}} \cdot \mathbf{d}^\alpha) {H}^\alpha$ is generally a different operator for different polarization choices of incoming and outgoing light given by $\boldsymbol{\epsilon}_{\textrm{in}}$ and $\boldsymbol{\epsilon}_{\textrm{out}}$, leading to many different spectra denoted $I_{\mu \nu, \mu'\nu'}(\omega)$.	
	Here we discuss the relations between Raman intensities that come from symmetry and the explicit form of the operators. 
	
	If we act on the system with a space group symmetry transformation, the Raman intensity should be the same. That is, the symmetries must act trivially on the Raman intensity. We first consider the action of the symmetries on the Raman operator. This corresponds to changing the polarization so that for a symmetry action parametrized by $\mathcal{O}_{\mu\mu'}$ we get
	\begin{align}\label{symms}
	R 
	&\to \sum_{\mu,\nu,\mu',\nu'} (\boldsymbol{\epsilon}_{\textrm{in}})_\mu \mathcal{O}_{\mu \mu'} R_{\mu' \nu'} \mathcal{O}_{\nu'\nu} (\boldsymbol{\epsilon}_{\textrm{out}})_{\nu}.
	\end{align}
	In particular $R_{\mu \nu} \to \mathcal{O}_{\mu \mu'} R_{\mu' \nu'} \mathcal{O}_{\nu'\nu}$. The action of the symmetries on the symmetric tensor $R_{\mu\nu}$ must form a representation of the space group, which is a quadratic representation since it has two spatial indices. 
The task is then to associate the operators in the $R_{\mu\nu}$ basis with the irreps. Then, since the symmetries must act trivially on $I$, the cross terms between $R_{\mu\nu}$ associated with different irreps must vanish.
	
	\subsection{3D lattices}
	
	For all $n<\infty$ the point group is $D_{2h}$. Since this is orthorhombic there are no symmetries mixing the $a,b,$ and $c$ directions so that none of the $R_{\mu\nu}$ with $\mu,\nu=a,b,c$ are mixed by the symmetry operations. This allows us to identify each of the 6 independent operators with an irrep.
The irreps of $D_{2h}$ can be found in standard tables. It has eight 1D irreps. However, since inversion can only act trivially on quadratic operators, $R$ cannot contain any component in four of them.   The four irreps that transform trivially under inversion are $A_{1g}, B_{1g}, B_{2g}$, and $B_{3g}$. 
	
Using Eq.~(\ref{symms}) one can check that the $aa$, $bb$, and $cc$ components of $R_{\mu\nu}$ transform trivially under every point group symmetry operation and therefore belong to the trivial $A_{1g}$ irrep. 
The other three components, $ab$, $ac$, and $bc$, transform under $B_{1g}, B_{2g}$, and $B_{3 g}$ respectively. In particular, two of the mirror symmetries act non-trivially in each of these irreps (e.g. $R_{ab} \xrightarrow{a\to -a} -R_{ab} $). To summarize, the symmetries imply that the only non-zero intensity cross terms are $I_{aa,bb}, I_{aa,cc},$ and $I_{bb,cc}$ and the other ones are zero (Recall Eq.~(\ref{I-terms})). 

Up to now we have only used the point group and thus these results hold for any Hamiltonian respecting the point group symmetries of the lattice. 
For a spin-exchange Hamiltonian we find further simplification upon writing out the Loudon-Fleury operator (Recall Eqs.~(\ref{Hex}) and~(\ref{R})).  
For the bond vectors (\ref{3Dds}) we get 
\begin{align} 
4 R_{aa} &= H^x_A + H^x_B + H^y_A + H^y_B \label{aa}\\
2 \sqrt{2} R_{ab} &= H^x_A - H^x_B + H^y_A - H^y_B \label{ab}\\
2 \sqrt{2} R_{bc} &= -H^x_A + H^x_B + H^y_A - H^y_B \label{bc}\\
4 R_{ac} &= -H^x_A - H^x_B + H^y_A + H^y_B \label{ac}\\
R_{cc} &= R_{aa}+H^z. \label{cc}
\end{align}
First we find $R_{bb} = 2R_{aa}$. This follows from the effective screw rotation discussed in the main text. 
Second, note that \mbox{$3 R_{aa} + R_{cc} = H$}. We call this relationship, which comes from the Loudon-Fleury form of the Raman operator, the LF relationship. 

Consider what this means for the correlation functions $I$. In the ground state $\braket{e^{i H t} R_{\mu\nu}e^{-i H t} \1}= 0$. Therefore, since \mbox{$R_{cc}\ket{0}=-3 R_{aa}\ket{0}+E\ket{0}$}, where $E$ is the energy of $\ket{0}$ we find that
\begin{align}
\braket{R_{\mu\nu}(t) R_{cc}(0)} = -3 \braket{R_{\mu\nu}(t) R_{aa}(0)}.
\end{align}
This gives $I_{cc} = 9 I_{aa} = -3 I_{aa,cc}$ for any nearest-neighbor spin-exchange Hamiltonian on these lattices. 

The combination of symmetries and relations imposed by the explicit form of the operators $R_{\mu\nu}$ reduce the number of independent, non-zero spectra to four. We choose to plot $I_{aa}$, $I_{ab}$, $I_{ac}$, and $I_{bc}$ as representatives. 

\subsection{2D Honeycomb}

The point group of the honeycomb lattice is $D_{3d}$,\cite{Sandilands} which has 3 irreps: $A_{1g}$, $A_{2g}$, and $E_g$. The $A_{1g}$ and $A_{2g}$ irreps are both 1D but $E_g$ is 2D because it involves nontrivial representation of $C_6$ rotation. In $A_{2g}$ each of the three reflection is represented by $-1$, which cannot be realized by a quadratic operator. We say that this channel is not Raman active.   

Unlike the case for the 3D lattices, the $C_6$ rotation symmetries do mix Raman operators in the $R_{\mu\nu}$ basis with $\mu,\nu = x,y$. In particular, the linear combination \mbox{$R_{A_{1g}} = \frac{R_{xx} + R_{yy}}{2}$} transforms under the single irrep $A_{1g}$ and the vector $(R_{xy},R_{E_g})$ transforms under the $E_g$ irrep, where \mbox{$R_{E_g} = \frac{R_{xx} - R_{yy}}{2}$}. One can check that the action of the symmetry elements on this vector realizes the full 2D representation of the $C_6$ rotations given in Eq.~(\ref{rep2}). 

Following the same procedure as for the 3D lattices we write out the Hamiltonian into its separate parts on each different bond (by orientation): \mbox{$H = \sum_{\alpha = x,y,z} H^\alpha $}. Then for any nearest-neighbor spin-exchange Hamiltonian respecting the lattice symmetry the LF Raman operators have the form
\begin{align}\label{RR2D}
R_{xx} &= \frac{3}{4} (H^x + H^y), 
 \hspace{.5cm}
& R_{yy} &= H^z + \frac{1}{3}R_{xx}, 
\\
R_{A_{1g}} &= H,
\hspace{.5cm}
& R_{E_g} &= -H^z + \frac{2}{3}R_{xx}, \\
R_{xy} &= 
\frac{\sqrt{3}}{4}(H^y-H^x).
\end{align}
We see that the $R_{A_{1g}}$ channel does not create any excitations since it is equal to the Hamiltonian. Therefore only the $E_g$ irrep is active on this lattice.

We can relate $R_{xy}$ and $R_{E_g}$ by considering the action of $C_6$ rotation on the correlation functions involving those operators. We find
\begin{align}
I_{xy,xy} \to \frac{1}{4} I_{xy,xy} + \frac{3}{4}I_{E_g,E_g} \pm \frac{\sqrt{3}}{2} I_{E_g,xy}.
\end{align}
This implies that $I_{E_g,xy}=0$ and that $I_{xy,xy} = I_{E_g,E_g}$.  

The fact that the Hamiltonian makes no excitations also implies that the operators $H^x + H^y$ and $-H^z$ give rise to the same correlation functions and, by some rearrangement, $R_{xx}$, $-R_{yy}$, and $R_{E_g}$ lead to the same Raman intensities. Finally, we have $I_{xx,xx} = I_{yy,yy} = -I_{xx,yy} = I_{xy,xy}$. Combining these relationships with the symmetry constraints we have $I_{xx,xx} = I_{yy,yy} = I_{xy,xy} = - I_{xx,yy}$ and $I_{xx,xy} = 0 = I_{yy,xy}$, which leaves only one independent spectrum. Considering an arbitrary polarization $\boldsymbol{\epsilon}_{\textrm{in}} = \cos \theta \hat{\mathbf{x}} + \sin \theta \hat{\mathbf{y}}$ and $\boldsymbol{\epsilon}_{\textrm{out}} = \cos \phi \hat{\mathbf{x}} + \sin \phi \hat{\mathbf{y}}$ we can compute $I$ in terms of $I_{xx,xx}$, for instance. In this case one finds that the angular dependence drops out and $I = I_{xx,xx}$ independent of polarization. 

When we take $J^x = J^y \ne J^z$ in the Kitaev Hamiltonian we break the $D_{3d}$ down to $D_{2h}$. In fact, this group remains the point group even if we let $J^x \ne J^y$. In this group the quadratic operators $R_{xx}$ and $R_{yy}$ act separately under the $A_{1g}$ irrep (they are even under reflection) and $R_{xy}$ falls under $B_{1g}$. The $B_{2g}$ and $B_{3g}$ irreps are not active because there are no bonds along the direction out of the honeycomb plane. The LF relationship still gives $I_{xx,xx} = I_{yy,yy} = -I_{xx,yy}$ and the symmetry gives $I_{xy,xx} = 0 = I_{xy,yy}$,  reducing $I$ to the sum of two independent correlation functions. We choose to plot $I_{xx}$ and $I_{xy}$. 

The inclusion of next-nearest-neighbors (NNNs) is quite simple under this formalism -- one simply treats these as different bonds over which $\alpha$ runs. For the honeycomb lattice there is no expected change in the polarization dependence because $R_{xx} + R_{yy} = H$ holds for any bond orientation. However, on the 3D lattices we find that the relation to the Hamimltonian for NNNs cannot be satisfied simultaneously with the NN one \mbox{$3 R_{aa} + R_{cc} = H$} because the coefficient $3$ in front of $R_{aa}$ results from the specific of the bond-orientations that occur at nearest-neighbor only. Therefore the differences in $I_{cc}$ and $9I_{aa}$, for instance, are a probe of terms beyond a NN spin-exchange Hamiltonian.

		\bibliographystyle{apsrev}
		\bibliography{References}

\begin{thebibliography}{97}
\bibitem[{GSf()}]{GSfoot}
\bibinfo{note}{In addition, depending on the boundary conditions and the flux
  configuration, the state with no quasiparticles may not respect the
  constraints $D_j = 1$.\cite{Pedrocchi} It was shown in Ref.
  \onlinecite{Pedrocchi} that in this case the physical ground state has one
  \ms in it. However, the corrections to the eigenvalues and even dynamic
  spectra are expected to be intensive and therefore vanish in the
  thermodynamic limit;\cite{Pedrocchi,Zschocke} we therefore work exclusively
  with the ground state of the extended Hilbert space.}
\bibitem[{ban()}]{bandfoot}
\bibinfo{note}{For the \Hs{0} lattice the eigenvalues can be computed
  analytically\cite{Lee2} since the four-site unit cell leads to a $4\times 4$
  matrix diagonalization problem.}
\bibitem[{wid()}]{widthfoot}
\bibinfo{note}{In Eq. (\ref{H}) we defined the Hamiltonian in terms of Pauli
  matrices. If we define the coupling in terms of spins, $H =
  \sum_{\left<ij\right>} \tilde{\Gamma}^{\alpha \beta}_{ij} S_i^\alpha
  S_j^\beta$, where $S_i^\alpha = \frac{1}{2} \sigma_i^\alpha$, the Raman
  bandwidth is $z\tilde{\Gamma}^{av}$ in place of $4z\Gamma^{av}$.}
\end{thebibliography}
		
	\end{document}